How Does Culture Evolve?

Liane Gabora

Note: There may be minor differences between this version and the final accepted version following page proofs.

Address for correspondence:
Department of Psychology, University of British Columbia
Fipke Centre for Innovative Research, 3247 University Way
Kelowna BC, V1V 1V7, CANADA
liane.gabora@ubc.ca

## Contents

**Introduction**

This chapter synthesizes a program of research devoted to developing a scientific framework for understanding how human culture evolves, using studies with human participants, computational models, and mathematical analyses. By ***culture***, I mean extrasomatic (external to the body) adaptations, including behaviour and artefacts, that are socially rather than genetically transmitted. Culture encompasses human endeavours ranging from poetry to architecture to animated cartoons. By ***evolution,*** I mean 'descent with modification,' i.e., cumulative, adaptive, open-ended change. Evolution is ***cumulative*** when new changes build on earlier ones. It is ***adaptive*** when innovations become increasingly fit—i.e., more beneficial in some way—to their bearers over time. It is ***open-ended*** when there is no *a priori* limitation to the innovations that arise. Because cultural change over time is cumulative, adaptive, and open-ended, culture—like biological species—evolves.

The evolution of culture could be said to ride piggyback on biological evolution, in the sense that it occurs in groups of humans, who owe their existence to biological evolution. However, culture can be studied independently of biology (much as one can study the behavior of whales with no special knowledge of the subatomic particles of which they are composed). Although we may meet only a fraction of the people who culturally influence us (by way of their inventions we use, their music we hear, and so forth), and we may meet only a fraction of the people who are affected by our *own* actions and ideas, we all, in some small way, contribute to this second evolutionary process, that of culture.

The research program focuses on a set of questions that are interrelated in that each step toward a solution to one of these questions provides fragmented glimpses of solutions to the others. The questions are:

1. **How did the *capacity* for creative culture evolve?** Although cultural *transmission* (in which one individual acquires elements of culture from another) is observed in many species, cultural *evolution* (in which elements of culture are not just transmitted but adaptively modified) is much rarer, and perhaps unique to our species. Research in psychology, neuroscience, anthropology, archaeology, and genetics, as well as mathematical and computational modelling, is increasingly converging toward a consistent picture of the cognitive and concomitant neural changes over the last few million years that enabled humans to, not just understand their world, but transform it.

2. **What *fuels* cultural innovation?** While the first branch of the research program focuses on historical, biologically evolved changes in the cognitive abilities underlying cultural evolution, this branch is focused on understanding, in a precise, rigorous way, the mechanisms underlying the creative processes by which cultural novelty is generated. In other words, it aims to understand the process by which existing knowledge and experience comes together to generate new technologies, songs, artworks, scientific theories, and so forth.

3. ***How* does human culture evolve?** A third branch of the research program seeks to understand how cultural evolution works, *i.e.,* does culture evolve, as do biological organisms, through natural selection, or by some other means? Are biological and cultural evolution isomorphic, *i.e.,* despite their superficial differences do they share a common algorithmic structure? By comparing and contrasting these two evolutionary processes, we gain a richer understanding of how evolutionary processes get started, and how they work.

This chapter consists of sections devoted to each of these three questions, interspersed with sections on computational models of cultural evolution, and cognitive autocatalytic networks, which possess the kind of self-organizing structure capable of evolving culture. The more one thinks about these three questions, the more it seemed they are not separate, but different threads of the same tangled web. The research program discussed here attempts to weave them into a tapestry of understanding. My research team is interdisciplinary, not because we particularly want to cross disciplines, but because that is where the questions take us. On the way to answering the above key questions, we stumble upon a host of other tantalizing questions:

- Why have humans transformed this planet in a way that no other species has?

- How does the creative process drive cumulative culture? How do new ideas build upon existing ones?

- Why do you feel so alive when immersed in a creative project? Why does life feel meaningful when you 'find your own voice', express 'how things look from where you stand', or take what you have learned and 'put your own spin on' it, especially when 'your spin' gets a stamp of approval from others?

The research program synthesizes research on cultural evolution, evolutionary theory, and creativity, using tools such as agent-bas ed models and autocatalytic networks (both of which are defined and discussed in Appendix A, and elaborated upon further in this chapter). What unites these projects is the aim of developing a psychologically, neurologically, and evolutionarily plausible framework for how ideas unfold over time; not just in the minds of individuals, but through direct and indirect interactions *amongst* individuals. This interdisciplinary undertaking has the potential to help us understand where we came from, who we are, and where we may be headed.

## Question 1: How did the *Capacity* for Creative Culture Evolve?

It has been proposed that humans possess two levels of complex, adaptive, self-organizing, evolving structures: a biological level, and a psychological level (Barton, 1994; Combs, 1996; Freeman, 1991; Pribram, 1994; Varela, Thompson, & Rosch, 1991). Of course, the second, psychological level—our minds—are *part of* living organisms, which are complex, adaptive, and self-organizing. However, the psychological level pertains not to cellular or organismal structures, but to the structures that steer our mental and cultural lives. I believe that the emergence of a new level of self-organizing structure was as critical to establishing cultural evolution as it was to the establishment of biological life. This suggests that the psychological level is responsible for humans' unique capacity for cumulative culture (Gabora, 1998, 2004).

In fact, this second level of self-organizing structure is not a *mind* exactly, but the mind *as experienced from the inside*: a *worldview*. While the term *mind* refers to the set of cognitive faculties including perception, judgement, memory, and consciousness, when we talk about a worldview, the emphasis is on a particular way of *seeing* the world and *being* in the world, including the basic outlook that guides how one navigates ones' unique experiences and understandings. It also includes the beliefs, attitudes, ideas, schemas, scripts, procedural and declarative knowledge, and habitual ways of thinking and acting. This second psychological level of self-organizing structure enables people to not just track their environment, learn from it, and interact with it in fixed ways, but adapt their responses to new situations, and reframe new knowledge in terms of what they already know. It enables people to detect inconsistencies, mull them over, and achieve more multifaceted understandings. As new information is assimilated into one's worldview, and the worldview changes to accommodate this information, it becomes more robust. Each idea or interpretation of a situation that someone comes up with is an expression of that person's worldview that can impact (albeit to perhaps a seemingly insignificant degree) how culture unfolds.

We now summarize research aimed at explaining the emergence of this second, cognitive level of self-organizing, integrated structure in humans, the level we consciously experience as a worldview. The research focuses on two cultural transitions as indicated by the archaeological record, for which there is converging anthropological and genetic evidence. Here we summarize this research in a nontechnical manner, and toward the end of the chapter, we outline computational and mathematical models of these cognitive transitions.

**The Origins of Cultural Evolution**

The earliest preserved signs of human culture include primitive stone tools devised over three million years ago (Lewis & Harmand, 2016). 1.76 million years ago (mya) there was a transition in tool technology from the Oldowan stone flake, made through simple repeated percussive action, to the Acheulian hand axe, which requires a skilled, multi-step process involving multiple different hierarchically organized actions (Edwards, 2001, Pargeter, Khreisheh, & Stout, 2019; Stout, Toth, Schick, & Chaminade, 2008). This was perhaps the earliest significant leap in cumulative culture. What caused it?

It has been suggested that the enlarged cranial capacity of *Homo erectus* enabled them to voluntarily retrieve and recursively modify memories and knowledge items independent of environmental cues (Donald 1991). Related hypotheses have been put forward by others (e.g., Hauser et al., 2002; Penn, Holyoak, & Povinelli, 2008). This capacity has been referred to as a self-triggered recall and rehearsal loop, or *representational redescription* (following Karmiloff-Smith (1992) who used the term in reference to developmental change). The capacity for representational redescription would have ushered forth a transition to a new mode of cognitive functioning. While *Homo habilis* was limited to the "here and now," representational redescription would have enabled *Homo erectus* to *chain* memories, thoughts, and actions into more complex ones, and progressively modify them, thereby gaining new perspectives on past or possible events, and even mime or re-enact them for others. However, this opens up another question: what made representational redescription possible?

A feasible answer can be found by examining the structure and dynamics of associative memory (Gabora, 2000b, 2010a). Neurons are sensitive to primitive stimulus attributes or 'microfeatures', such as sounds of a particular pitch or lines of a particular orientation. Experiences encoded in memory are *distributed* across cell assemblies of neurons, and each neuron participates in the encoding of many experiences. Memory is also *content-addressable:* similar stimuli activate overlapping distributions of neurons.

With larger brains, experiences could be encoded in more detail, enabling a transition from coarse-grained to fine-grained memory. Fine-grained memory enabled more distinctions to be made, so concepts and ideas could be encoded in more detail. As a simplified example, **Figure 1** shows a particular experience encoded in terms of only two properties in *Homo habilis*, while the same experience is encoded in terms of three properties in *Homo erectus.* At the level of neural cell assemblies, more fine-grained mental representations meant that more feature-specific neurons participated in the encoding of any given memory or knowledge item, which in turn meant more possible routes by which they could evoke remindings of one another and interact, and more ways of reframing or redescribing current experiences in terms of past experiences. This paved the way for streams of abstract thought; ideas could now be reprocessed until they cohered with existing conceptual structure. At this point, it was possible to think about an idea in relation to other closely related ideas, and thereby forge clusters of mutually consistent ideas, which allowed for a narrow kind of creativity, limited to minor adaptations of existing ideas. However, the mind was not yet highly integrated, nor truly creative; it could not yet forge connections between seemingly disparate ideas, as in the formation of analogies.

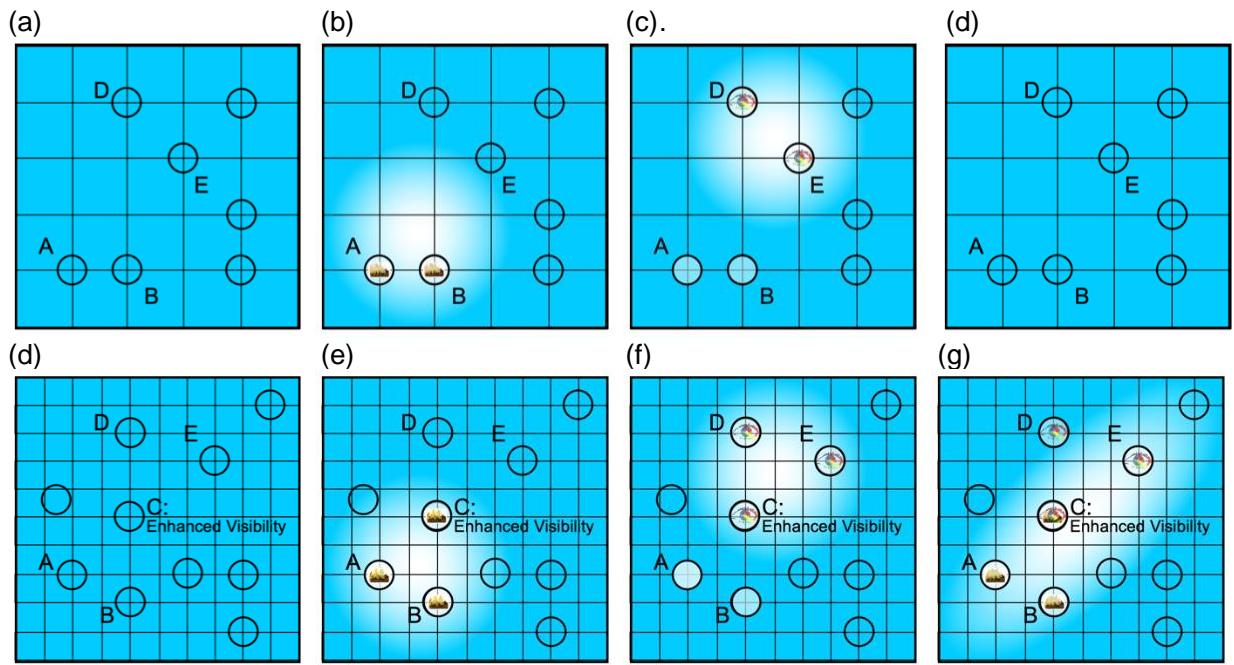

**Figure 1**. Upper row: Coarse-grained memory of *Homo habilis*. (a) Each black ring represents a neuron tuned to respond to a specific feature of the world. (b) The presence of fire activates a distributed mental representation of fire, represented here by neurons A and B, which respond to features of fire (such as 'hot' or 'bright'). (c) The need to see while traveling in the dark activates a distributed mental representation of this experience, represented here by neurons D and E. (d) The need to see while traveling in the dark does not activate the thought of fire in the coarse-grained memory of *Homo habilis*, because the mental representations do not overlap. Lower row: Fine-grained memory of *Homo erectus*. The density of neurons is higher, and includes a neuron that specifically responds to the feature 'increased visibility'. (e) The presence of fire activates a distributed mental representation of fire, encoded by not just neurons A and B, but also C, since fire enhances visibility. (f) The need to see while traveling in the dark activates a distributed mental representation of this experience, represented here by not just neurons D, E, but also C, since there is a perceived need for enhanced visibility. (g) Neuron C acts as a gateway to self-triggered recall of the fire, which also activates neuron C, thereby generating the idea of taking a stick from the fire and using it to light the way while traveling in the dark. This insight was possible for the fine-grained memory of *Homo erectus*, but not for the coarse-grained memory of *Homo habilis*.

Note that this explanation of insight at the level of neural cell assemblies is consistent with research on *structured imagination*, which shows that the emergence of new ideas through concept combination is guided by the structure of existing categories and knowledge domains, which use contextual information to influence the flow of activation, and identify connections (Ward, 1994; Ward et al., 2002).

**The Origins of Behavioral and Cognitive Modernity[1]**
Between 100,000 and 30,000 years ago, in the Middle Upper Paleolithic, there was a proliferation in cultural artifacts of both utilitarian and aesthetic value, including a variety of complex, task-specific tools (Ambrose, 2001), representational art (Aubert et al, 2018; Nelson, 2008; Pike et al., 2012), personal symbolic ornamentation (D'Errico, 2009), complex living spaces (Otte, 2012), sophisticated ways of obtaining food (Erlandson, 2001), burial sites indicating ritual (Hovers, Lani, Bar-Yosef, & Vandermeersch, 2003), and possibly religion (Rappaport, 1989), and syntactically rich language (Bickerton, 2007).

Psychological explanations for this cultural transition have been proposed (summarized in Gabora & Smith, 2018). Some emphasize the onset of enhanced social skills (Tomasello, 2014), tool use (Osiurak & Reynaud, 2020), or language (Corballis, 2011). However, the fact that such diverse abilities and artifacts came about at this time suggests that it was due to the onset of a more general cognitive capacity, one that underwrote social skills, tool use, *and* language. Accordingly, it was proposed that this cultural transition was due to onset of the ability to, not just recursively redescribe mental contents, but reflect on ideas from different perspectives, at different levels of abstraction, using different modes of thought (Gabora, 2003, 2008a; Winslow & Gabora, 2022). To contribute to cultural change, different knowledge domains must be accessible to one another, so as to combine concepts, or adapt existing ideas and methodologies to new situations or preferences. However, if a worldview were *too* integrated, everything would be reminiscent of everything else, which could interfere with routine survival tasks. While divergent thought is helpful for integrating information and generating ideas when stuck, in day-to-day living it could be disruptive. Therefore, the onset of divergent thought had to be accompanied by the capacity to reign it in.

The ability to shift between different modes of thought—an explicit, convergent, *analytic mode* conducive to logical problem solving, and an implicit, divergent, *associative mode* conducive to insight and breaking out of a rut—is referred to as *contextual focus* (CF) (Gabora, 2003).[2] CF enabled our ancestors to tailor their mode of thought to the demands of the situation, such that the fruits of one mode of thought could be used as ingredients for another, resulting in a richer understanding of the world, and enhanced potential to modify it. By shifting between modes of thought, our ancestors could not just envision radical new ideas, but test and refine them, and carry them through to completion. By bridging concepts and ideas from different domains and arenas of life, and refining these connections by looking at them from different perspectives, and at multiple hierarchical levels of abstraction, the mind became increasingly web-like. Eventually, a particular individual's mental representations crossed a percolation threshold and underwent a phase transition to a new state in which they were sufficiently integrated and mutually accessible that they collectively formed the second level of self-organizing structure discussed earlier: the psychological level. This individual was now capable of the telltale mental acts of an integrated cognitive network, such as metaphor, analogy, and combining concepts or ideas from different domains. Moreover, by sharing fragments of this integrated mental structure with others in the form of stories, and prompting others to ask questions and make connections, this individual could potentially stimulate this kind of phase transition in the minds of others.

---

[1] There is some debate as to whether the archaeological record reflects a genuine transition, since claims to this effect are based on the European Paleolithic record, and largely exclude the lesser-known African record (Fisher & Ridley, 2013; McBrearty & Brooks, 2000).

[2] CF is reminiscent of dual processing theories (e.g., Evans, 2003). However, while dual processing theories generally attribute abstract, counterfactual thinking solely to a more recently evolved 'deliberate' mode, according to the CF hypothesis, abstract, hypothetical thought is possible in either, but it differs in the two modes, with divergent thought being conducive to flights of fancy, and convergent thought being conducive to logical analysis (Sowden, Pringle, & Gabora, 2014).

It was further proposed that CF was made possible by genetic mutation, possibly involving FOXP2, which is known to have undergone human-specific mutations in the Paleolithic era (Chrusch & Gabora, 2014; Gabora & Smith, 2018). FOXP2, once thought to be the 'language gene,' is not uniquely associated with language. The idea is that, in its modern form, FOXP2 (or some other gene) enabled fine-tuning of the neurological mechanisms underlying the capacity to shift between processing modes by varying the size of the activated region of memory.

The proposal, then, is that the origin of culture was due to the onset of representational redescription, enabling interaction between closely related memories and knowledge. Much later, the origins of behavioral and cognitive modernity was due to the onset of CF, enabling both the bridging of distant knowledge domains, and the refinement of knowledge at multiple hierarchical levels of abstraction. In the sections to come, we will examine how this proposal was tested and refined using computational and mathematical models.

**Computational Modeling of Transitions in Human Cognition**
It is difficult to experimentally test hypotheses about how the creative abilities underlying cultural transitions evolved. ***Agent-based modeling*** is a computational methodology in which artificial agents are used to represent interacting individuals. It enables us to address questions about the workings of collectives such as societies. It is particularly valuable for answering questions that lie at the interface between anthropology and psychology, owing both to (1) the difficulty of experimentally manipulating a variable (such as the average amount by which one invention differs from its predecessor) and observing its impact on cumulative culture over time, and (2) the sparseness of the pre-modern archaeological record. Although methods for analyzing these remains are becoming increasingly sophisticated, they cannot always distinguish amongst different theories.

EVOC (for EVOlution of Culture) and its predecessor MAV (for Meme And Variations, a pun on the 'theme and variations' musical form) is an agent-based model of cultural evolution that consists of neural network-based agents that invent new actions and imitate neighbors' actions in a two dimensional grid-cell world (Gabora, 1995, 2008b; Gabora & Tseng, 2017; Gabora & Smith, 2018). The core of each agent is a neural network in which mental representations of body parts (e.g., left arm) are encoded by input and output nodes, and more abstract notions (e.g. arm, degree of symmetry, or activity level) are encoded by hidden nodes.[3]

In each iteration of a run of EVOC, each agent decides probabilistically whether to (1) invent a new action, by modifying one an action currently in their repertoire, or, (2) if one of their neighbours is implementing an action than is fitter than their own current action, they can learn the action their neigbhour is implementing by imitating it. Over the course of a run, agents learn generalizations concerning what kinds of actions are useful, or have a high "fitness", with respect to a particular goal, and use this acquired knowledge to modify ideas for actions before transmitting them to other agents. The effectiveness, or 'fitness' of their actions are assessed by one or more predefined fitness functions, which evaluate how well their current behavior fulfills their current needs. They can have either one need or many, each of which is fulfilled by differect actions. The agents can be as sparsely or densely populated as the modeler wishes, and their world can be with or without borders, which can be either permeable or impermeable (in which agents on one side of a border cannot imitate agents on the other side). A model such as this is a vast simplification, and results obtained with it may or may not have direct bearing on complex human societies, but it allows us to vary one parameter while holding others constant, and thereby test hypotheses that could otherwise not be tested. It provides new ways to think about and understand what is going on.

---

[3] Learning and training of the neural networks is described elsewhere (Gabora, 2008; Gabora & Tseng, 2017; Gabora & Smith, 2018).

Parameters that can be turned on or off include *knowledge based-operators:* the capacity to bias innovation using trends observed over the course of a run about what makes a good action (for example, if actions that involve symmetrical movement of the arms tend to have high fitness, new actions are more likely to entail symmetrical arm movement). One can also turn on or off the capacity for *mental stimulation:* the ability to assess the fitness of a neighbour's action before imitating and implementing it. The user can also vary the percentage of agents capable of inventing, the mean ratio of inventing to imitating across the artificial society, and the capacity for agents to adjust how creative they are based on the success of past inventions.

EVOC and MAV exhibit typical evolutionary patterns, such as a cumulative increase in the fitness (effectiveness) and complexity of cultural outputs over time, as shown in **Figure 2a**. If even a small fraction of agents are creative, new ideas spread in waves by imitation throughout the society, reaching other creators who put another spin on them. Thus, over time, they accumulate change, and evolve. Another typical evolutionary pattern exhibited by these computational models of cultural evolution is an increase in diversity as the space of possibilities is explored followed by a decrease as agents converge on the fittest possibilities, as shown in **Figure 2b**. EVOC has been also used to model how the mean fitness and diversity of cultural elements is affected by factors such as population size and density, and borders that affect transmission between populations.

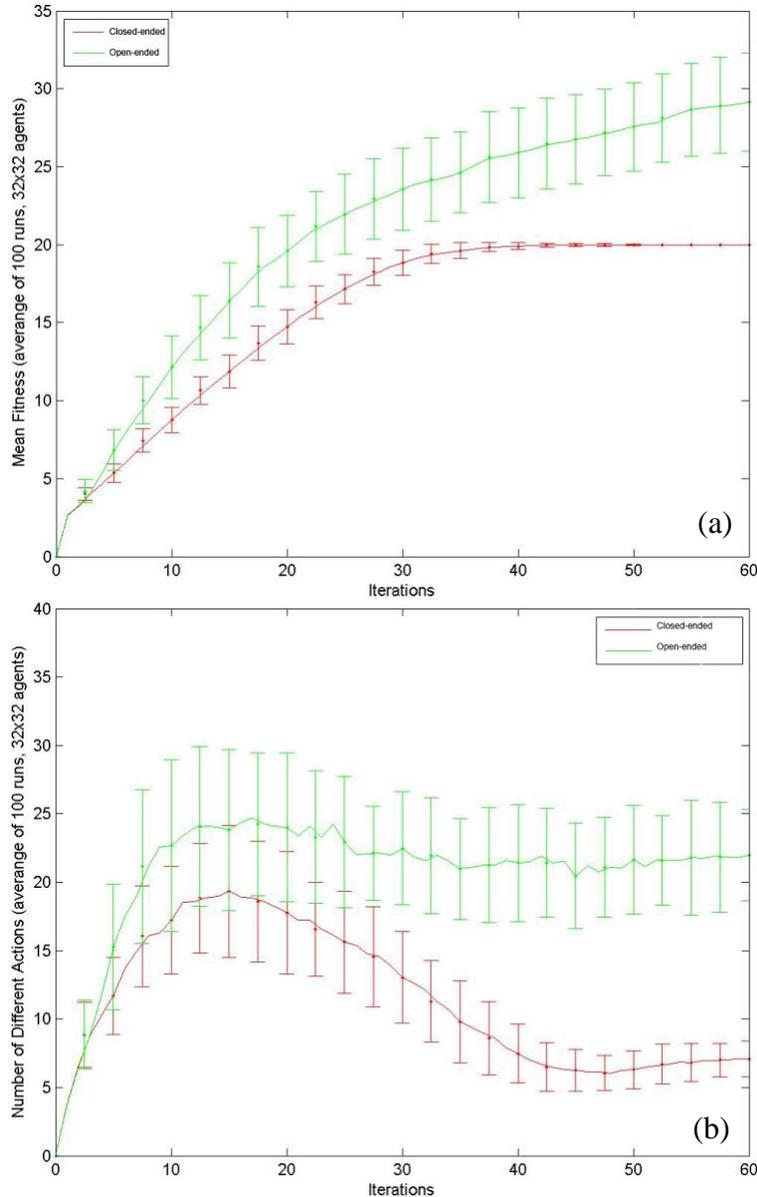

**Figure 2**. Results obtained using EVOC, a computational model of cultural evolution. In 'open-ended' runs, agents could generate fitter, increasingly complex and outputs, whereas in 'closed-ended,' once the maximally fit output was obtained, no further improvement was possible. (a) The top graph shows how the fitness of cultural outputs increases over time. (b) The bottom graph shows how the diversity of cultural outputs first increases as the space of possible variants is explored, and then decreases, as the artificial society converges on the fittest cultural outputs. (From Gabora, 2018b.)

*Computational Model of Origin of Cultural Evolution*
EVOC has been used to address the first question we asked in the introduction: How did the capacity for cultural evolution come about? In other words, what changes came about in our ancestors due to *biological evolution* that enabled complex *culture* to exist?

     Recall the hypothesis that cultural evolution was made possible by onset of the capacity self-triggered recall or representational redescription, which enabled the chaining and progressive modification of thoughts and actions. This was tested in EVOC by comparing runs in

which agents were limited to single-step actions to runs in which they could chain ideas together to generate multi-step actions (Gabora, Chia, & Firouzi, 2013; Gabora & Smith, 2018). Chaining increased both the mean fitness of cultural outputs across the artificial society, as illustrated in **Figure 3a** (top graph), and the diversity of cultural outputs, as illustrated in **Figure 3b** (bottom graph). While chaining and no-chaining runs both converged on optimal actions, without chaining this set was static, but with chaining it was in constant flux, as ever-fitter actions were found. While without chaining there was a ceiling on the mean fitness of actions, with chaining there was no such ceiling, and chaining also enhanced the effectiveness of the ability to learn trends (not shown here; see (Gabora & Smith, 2018) for details). These findings support the hypothesis that the ability to chain ideas together *can* transform a culturally static society into one characterized by open-ended novelty; thus, it strengthens the feasibility of our theory as an explanation for the origin of cumulative culture.

*Computational Model of Onset of Behavioral and Cognitive Modernity*
The hypothesis that contextual focus (CF) enabled behavioral and cognitive modernity and cumulative cultural evolution was also tested using computational models (e.g., Gabora & DiPaola, 2012), including EVOC (Gabora, Chia, & Firouzi, 2013; Gabora & Smith, 2018). If the fitness of an agent's outputs was low, it would temporarily shift to a more divergent mode by increasing α: the degree to which a newly invented idea deviates from the idea on which it was based. Conversely, If an agent's outputs were already very adapted to their artificial environment, it would shift to a more convergent processing mode by decreasing α.

The results were just as we had hypothesized: the mean fitness of actions across the society increased with CF, as illustrated in **Figure 3a**, as did the diversity, as illustrated in **Figure 3b**. Moreover, CF was particularly effective when the fitness function changed at iteration 50 (as indicated in **Figure 3a** by the flattening of the CF and 'Chaining + CF' curves just prior to iteration 50, and the marked increase in the slope of these curves just after iteration 50). This supports CF's hypothesized role in breaking out of a rut and adapting to new or changing environments.

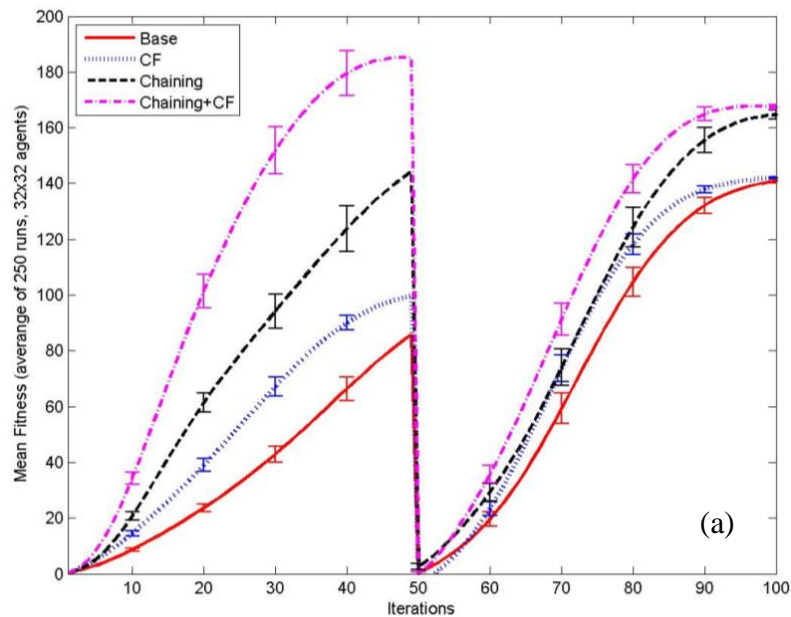

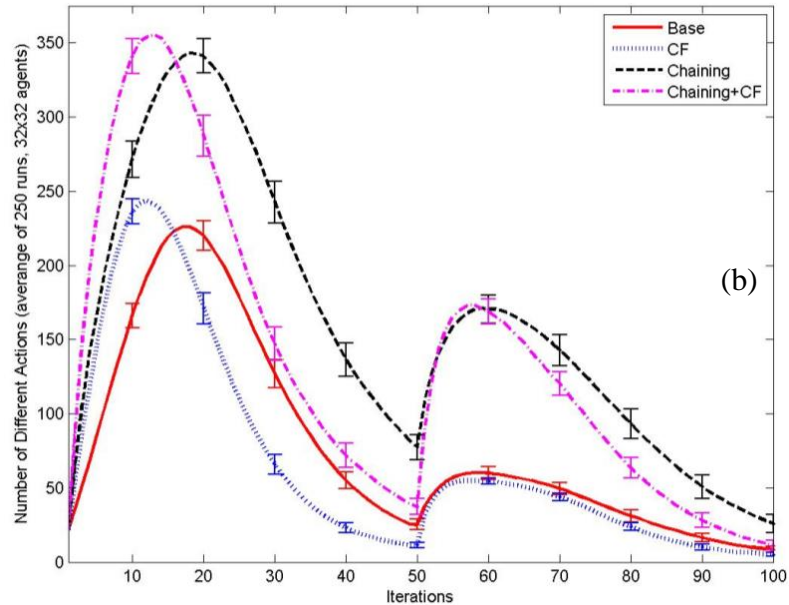

**Figure 3**. Mean fitness (graph (a), on top) and diversity (graph (b), on the bottom) of cultural outputs across the artificial EVOC society with both chaining and CF, chaining only, CF only, and neither chaining nor CF. Data are means of 500 runs. (From Gabora & Smith, 2018.)

These findings show how a cultural evolutionary framework can provide a valuable perspective on creativity. Note, however, that although chaining made the variety of novel outputs open-ended, and this became even more pronounced with CF, these novel outputs were nonetheless predictable. Chaining and CF did not open up new cultural niches in the sense that, for example, the invention of cars created niches for the invention of things like seatbelts and stoplights. EVOC in its current form could not solve *insight problems*, which require restructuring the solution space. Nonetheless it is sufficient to illustrate the effectiveness of chaining and CF. Building on a related research program in concept combination (e.g., Aerts, Gabora, & Sozzo, 2013), models of concepts provide further support, showing that CF is conducive to making creative connections by placing concepts in new contexts (Gabora & Aerts 2009; Kitto, Bruza, & Gabora, 2012; Veloz, Gabora, Eyjolfson, & Aerts, 2011).

**Question 2: What *Fuels* Cultural Innovation?**

Much research on cultural evolution is squarely focused on the social transmission of cultural knowledge. While lip service is paid to the creative processes by which this knowledge comes about in the first place, in empirical, computational, and mathematical models, novelty is often attributed to trivial processes such as 'cultural mutation' (Kandler, Wilder, & Fortunato, 2017) or 'copying error' (Schillinger, Mesoudi, Lycett, 2016), but clearly there is much more to it than that. We cannot understand how culture evolves without understanding the strategic, intuitive, creative processes that generate cultural novelty.

This section begins by outlining experiments with EVOC designed to understand the relationship between creativity and cultural evolution. We then discuss a theory of creativity that explicitly grew out of early research on cultural evolution, referred to as the honing theory of creativity, and examine empirical support for it.

**The Impact of Leadership and Social Media on Creative Cultural Evolution**
Throughout history, there have been leaders who were imitated more frequently than the common person. An interesting question is: what is the impact of leadership on creativity and the evolution of culture?

This question was also investigated in the EVOC agent-based model discussed above, using the *broadcasting* function. With broadcasting turned on, the action of a leader is visible to, not just immediate neighbors, but all agents, thereby simulating the effects of public performances, television, or social media, on patterns of cultural change. When broadcasting is turned on, each agent adds the broadcaster as a possible source of imitable actions. One or more agents are randomly chosen as broadcasters before the run, or alternatively, the user can specify that each iteration, the agent with the fittest action serves as the broadcaster.

When the user selects a broadcaster at random, broadcasting has no effect on the fitness of cultural outputs, but when the user chooses to have the broadcaster be the agent with the fittest actions, there is a modest increase in the fitness of actions. In either case, broadcasting accelerates convergence on optimal actions, thereby consistently reducing diversity. This can be seen in **Figure 4** by comparing column 1 with no leader (i.e., 0 broadcasters) to column 2 with one leader (a single broadcaster). Here, we see the previously mentioned general evolutionary trend of an increase in diversity as the space of possibilities is explored, followed by decreasing diversity as agents converge on the fittest possibilities, but we also see that leadership accentuates this normal decrease in diversity, since everyone starts to copy what the leader is doing. The total number of different actions after 20 iterations decreases from eight to five when a leader is introduced, and the percentage of agents executing the most popular action increases from 41\% to 84\%. This result appears to be consistent with evolutionary game theoretic results, obtained both analytically and with agent-based simulations, which suggest that the degree to which people explore new strategies is inversely related to norm strength, cultural inertia, and the need for social coordination (De, Nau & Gelfand, 2017). In human societies, leaders may bring about a decrease in diversity through norm enforcement and social coordination.

**Figure 4** also shows the impact of a dictatorial style of leadership (one broadcaster) versus a more distributed style of leadership (multiple broadcasters). As we go from one broadcaster to five, the total number of different actions after 20 iterations increases from five to nine, and the percentage of agents executing the most popular action decreases from 84% to 31%. Thus, the leadership-induced decrease in diversity appears to be mitigated by a more distributed form of leadership.

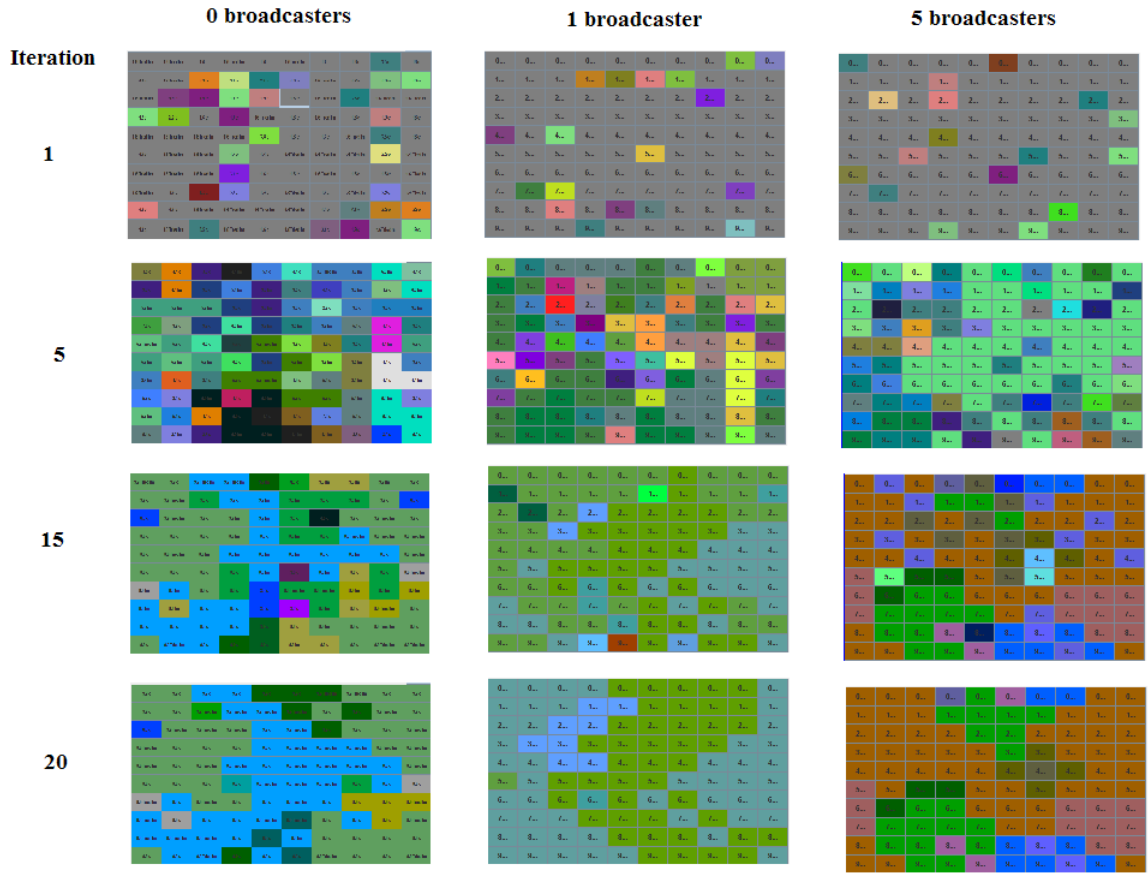

**Figure 4.** Diversity of actions over a run with 0, 1, and 5 broadcasters. Different actions are represented by differently coloured cells. In each case, there is an increase followed by a decrease in diversity over time (moving down any column from the first iteration at the top, to the 20th iteration at the bottom) but this becomes less pronounced with additional leaders. In these experiments, broadcasters were chosen at random every iteration, and when there were multiple broadcasters, agents imitated the broadcaster whose action was most similar to their own. (Adapted from Gabora, 2008b.)

## Balancing Creativity with Continuity

Cultural evolution, like any evolutionary process, requires a balance of change and continuity. *Change* injects new, potentially fitter variants into the process, and *continuity* helps guard against the loss of fit variants. The goal of next set of experiments was to better understand how cultural evolution balances creative change with continuity.

### *Varying the Ratio of Inventing to Imitating*

To investigate the optimal balance of inventing to imitating, the invention-to-imitation ratio was systematically varied from 0 to 1. When agents never invented, there was nothing to imitate, and there was no cultural evolution at all. If the ratio of invention to imitation was even marginally greater than 0, cumulative cultural evolution was suboptimal, but it was possible; the mean fitness of cultural outputs increased over time. Moreover, all agents converged on optimal cultural outputs, i.e., they all eventually implemented the fittest currently possible actions. When all agents always invented and never imitated, cumulative cultural evolution was again possible, yet suboptimal, because fit ideas were not dispersing through the society. The society as a whole performed optimally when there was a mixture of inventing and imitating, with the optimal ratio of the two being approximately 1:1.[4] This showed that, as in biological evolution, culture evolves most efficiently when the novelty-generating process of creativity is tempered with the continuity-fostering process of imitation.

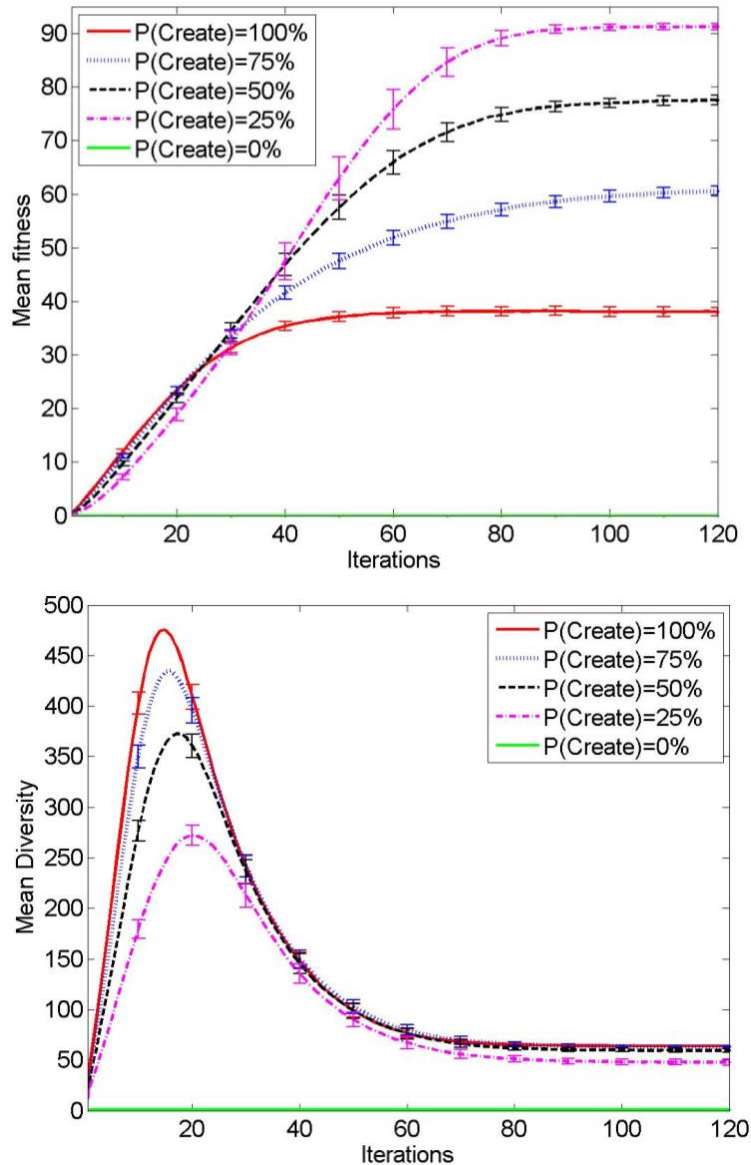

**Figure 5**. Fitness (top) and diversity (bottom) of cultural outputs with different ratios of inventing to imitating. (From Gabora, 2018).

In human societies, though creative individuals generate the novelty that fuels cultural evolution, absorption in their creative process impedes the diffusion of proven solutions, effectively rupturing the fabric of the artificial society. Thus, as hypothesized, in EVOC, a balance between creative (novelty-injecting) and conformist (continuity-maintaining) agents did indeed ensure that new ideas came about and, if effective, were not easily lost by society as a whole.

---

[4] The exact value of the optimal invention to imitation ratio depends on the difficulty of the fitness function; for example, with the difficult fitness function shown in **Figure 5**, it was significantly lower than 1:1. A fitness function is difficult when there are local maxima (i.e., when there is a tendency to get stuck on local suboptimal solutions), or when there is cultural epistasis (e.g., in EVOC, cultural epistasis is present if the value of a particular kind of movement for one body part depends on what one or more other body parts are doing).

*Varying the Ratio of Creators to Conformers*

The finding that very high levels of creativity can be detrimental for society led to the hypothesis that there is an adaptive value to society's ambivalent attitude toward creativity; society as a whole may benefit by being composed of both a conventional workforce and what Florida (2002) refers to as the 'creative class.' This was investigated by introducing two types of agents: *conformers* that only obtained new actions through imitation, and *creators* that obtained new actions either through imitation *or* invention (Gabora & Firouzi, 2012; Gabora & Smith, 2018). A given agent was either a creator or an imitator throughout the run, and whether that creator invented or imitated in a given iteration fluctuated stochastically. We systematically varied $C$, the proportion of creators to imitators in the artificial society, and $p$, how creative the creators were. As illustrated in **Figure 6**, there was a trade-off between $C$ and $p$ thus providing further evidence that society, as a whole, functions optimally when creativity is tempered with continuity.

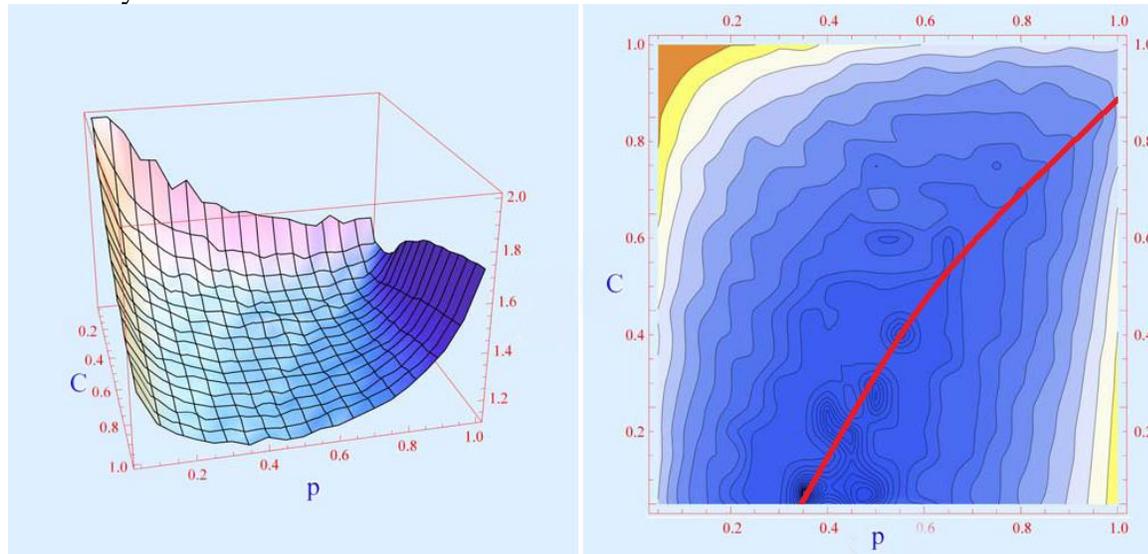

**Figure 6**. The effect of varying the percentage of creators, $C$, and how creative they are, $p$, on the mean fitness of ideas in EVOC. 3D graph (left) and contour plot (right) for the average mean fitness for different values of $C$ and $p$ using discounting to ensure that the present value of any given benefit with respect to idea fitness diminishes as a function of elapsed time before that benefit is realized. The z-axis is reversed to obtain an unobstructed view of the surface; therefore, lower values indicate higher mean fitness. The red line in the contour plot shows the position of a clear ridge in the fitness landscape, which indicates that the optimal values of $C$ and $p$ that are sub-maximal (i.e., lower than their highest possible value) for most {$C$, $p$} settings. In other words, it illustrates a tradeoff between how many creators there are, and how creative they should be. The same pattern of results was obtained analyzing just one point in time and using a different discounting method (not shown). (Adapted from Gabora, & Firouzi, 2012.)

These results suggest that excess creativity at the individual level can be detrimental at the level of the society because creators invest in unproven ideas at the expense of propagating proven ideas. They are like localized 'rips' in the 'fabric of society,' impeding the spread of known solutions. From this perspective, it perhaps makes sense that before creative people are embraced they must first prove themselves.

*Social Regulation of How Creative People Are*

We then wondered if, in real societies, individuals adjust how creative they are in accordance with their perceived creative success. This might happen, for example, by encouraging successful creators, and ignoring or ostracizing those whose inventions are deemed unfit (or are ahead of their time).

A first step in investigating this was to determine whether it is algorithmically possible to increase the mean fitness of ideas in a society by enabling them to self-regulate how creative they are (Gabora & Tseng, 2017). To test the hypothesis that the mean fitness of cultural outputs across society increases faster with social regulation (SR) than without it, we increased the relative frequency of invention for agents that generated superior ideas, and decreased it for agents that generated inferior ideas. Thus, when SR was on, if the relative fitness of a particular agent's ideas was high, that agent invented more often, and if it, was low that agent imitated more often. *p(C)* was initialized at 0.5 for both SR and non-SR societies.

When social regulation was introduced into the artificial society, the mean fitness of the cultural outputs was higher, as shown in **Figure 7**. The typical pattern was observed with respect to the number of different ideas across the population: an increase as the space of possibilities is explored, followed by a decrease as agents converge on fit actions. However, this pattern occurred earlier, and was more pronounced, in societies with SR than societies without it. Thus, societies in which the bearers of successful ideas have more opportunity to create, and the bearers of unsuccessful ideas have fewer opportunities to create, tend to be fitter, to generate more different ideas, and to converge on good ideas more quickly.

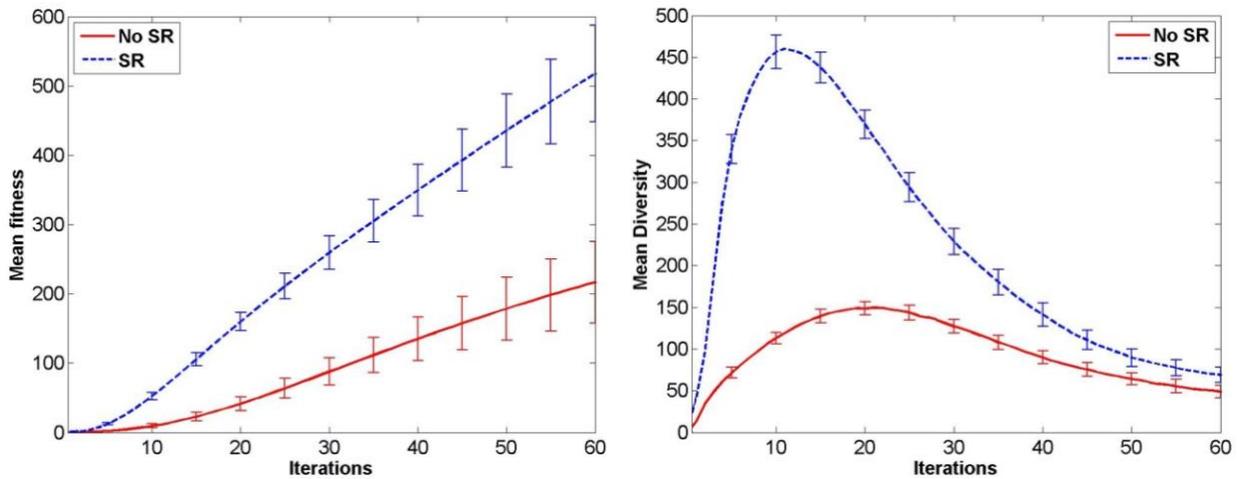

**Figure 7**. Mean fitness (left) and diversity (right) of cultural outputs across all agents over the duration of the run with, and without, social regulation. Data are averages across 250 runs. (From Gabora & Tseng, 2014).

As illustrated in **Figure 8**, societies with SR ended up segregating into two distinct groups: one that primarily invented, and one that primarily imitated. This means that the observed increase in fitness can indeed be attributed to increasingly pronounced individual differences in their degree of creative expression over the course of a run. Agents that generated superior cultural outputs had more opportunity to do so, while agents that generated inferior cultural outputs became more likely to propagate proven effective ideas rather than reinvent the wheel. Thus, we see the emergence of Florida's 'creative class'.

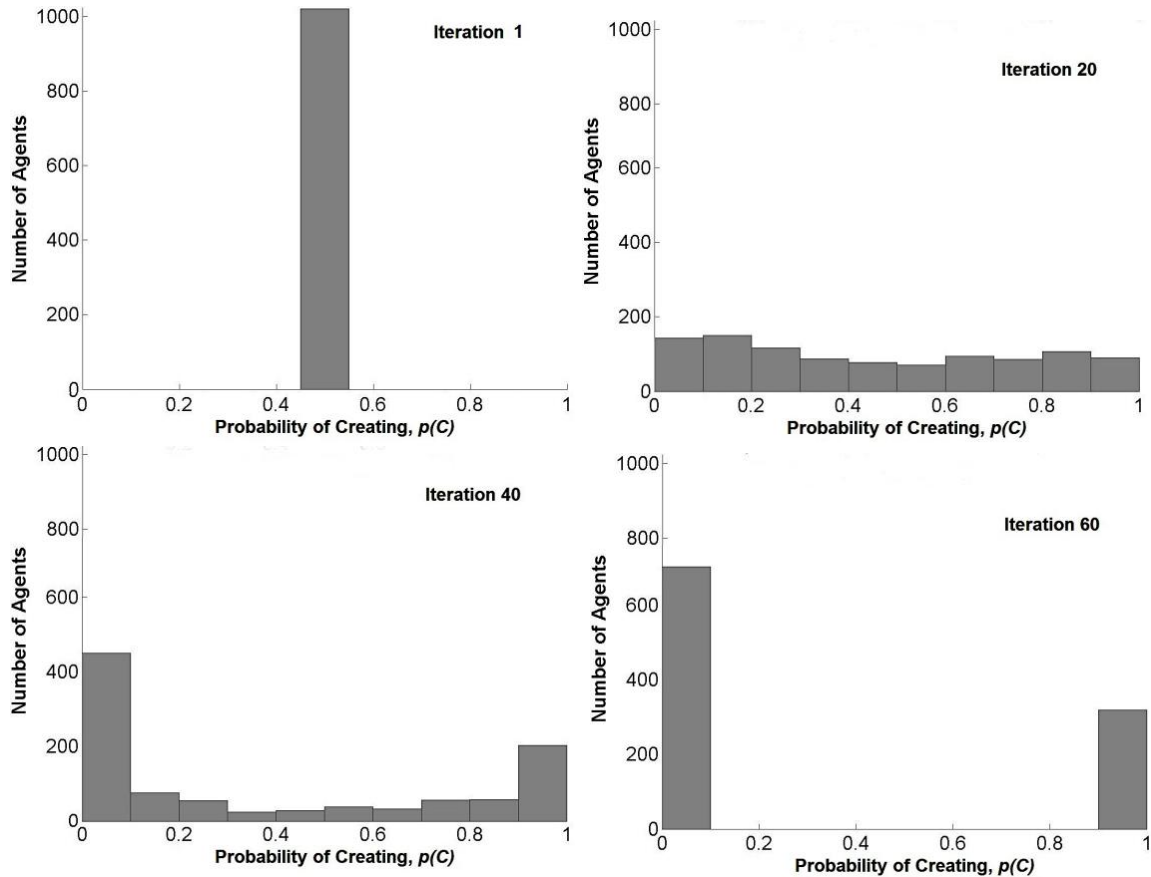

**Figure 8**. At the beginning of the run all agents created and imitated with equal probability. Midway through a run their *p(C)* values were distributed along the range from 0 to 1. By iteration 60 they had segregated into two distinct groups: conformers (with *p(C)* from 0 to 0.1) and creators (with *p(C)* from 0.9 to 1). (From Gabora & Tseng, 2014).

### Creativity: The Process that Drives Cultural Change

Though we have gained much in the way of fragmentary knowledge about creativity, an integrated framework for creativity eludes us (Bowden, Jung-Beeman, Fleck, & Kounios, 2005; Kaufman & Sternberg, 2010). Because *creativity* is such a multidisciplinary construct, it has been defined and studied in different ways. Some approaches emphasize novelty and originality, while others emphasize value and appropriateness, and they differ too with respect to their emphasis on individuals versus groups, and on ideas versus outputs, or social recognition of those outputs (Kaufman & Glăveanu, 2019).

In my view, placing the emphasis on outputs is problematic. If Person A comes up with an idea in isolation while Person B arrives at the same idea by slightly tweaking Person C's near-miss of the idea, surely Person A is more creative than Person B, though that isn't evident from the outputs. I view *creativity* as the cognitive process by which an intrinsically-motivated individual restructures his or her worldview, often (but not necessarily) culminating in the execution of an idea, and its debut appearance in the creator's cultural milieu. In short, I view creativity as what enables minds to self-organize, and thereby culture to evolve.

My realization that understanding creativity is vital to making headway in cultural evolution research dates back to my work as a graduate student on the first version of my computational model of cultural evolution (Gabora, 1995). It didn't seem like the artificial agents' 'creativity algorithm' had much to do with what happens in people when they are immersed in a creative task. I figured I'd have to learn a little psychology to get that part right. Creativity is perplexing; decades later, I still haven't got it entirely figured out. However, I believe I have it sufficiently worked out to set forth, in broad strokes, an integrated framework for addressing the interrelated questions of how culture evolves, and how the creative process works.

Synthesizing research in neuroscience, cognitive science, and machine learning, I developed the *honing theory of creativity* (Gabora, 2017). Since some of the core tenets of the theory may seem counterintuitive at first, we briefly examine the research that led to this theory before laying it out.

### Concept Combination: The Heart of the Creative Process

All creativity has, at its core, an act of seeing a concept from a new context (*e.g.,* in the context PLAYGROUND you might conceive of a TIRE as a SWING), or combining concepts (e.g., combining BEANBAG and CHAIR to invent BEANBAG CHAIR). To understand creativity, we need a formal theory of how concepts interact. However, people use conjunctions and disjunctions of concepts in ways that violate classical logic (Hampton, 1987, 1988; Osherson & Smith, 1981) (*e.g.*, people do not rate GUPPY as a typical exemplar of PET, nor of FISH, but they rate it as a typical exemplar of PET FISH). This makes concepts resistant to mathematical description, a problem that has been called the greatest challenge facing cognitive science (Fodor, 1998), due to its profound implications for language and cognition.

I knew that physicists had encountered similar difficulties describing *interactions* (in their case, interactions amongst particles), and to describe these interactions they devised a new kind of mathematics. One phenomenon they encountered was the *observer effect:* the act of observing a system unavoidably disturbs it, affecting its state. It seemed that something similar went on with respect to concepts; the context that causes a concept to come to mind unavoidably colours your experience of that concept. Other phenomena that I thought might possibly be useful for explaining what happens when concepts combine are entanglement and interference. *Entanglement* is a phenomenon first encountered in particle physics wherein the state of one entity cannot be described independently of the state of another, and any measurement performed on one influences the other. *Interference* is the annihilation of the crest of one wave by the trough of another when they interact. I convinced my PhD supervisor that concepts might also exhibit these effects, and that their behavior could be modelled using a generalization of the mathematics originally developed for quantum mechanics. We wrote the first papers on this (Aerts & Gabora, 2005a,b; Gabora & Aerts, 2002), which contributed to the birth of a field that has (somewhat unfortunately) come to be called *quantum cognition*. The approach has successfully modeled many aspects of cognition that eluded conventional approaches (Wang, Busemeyer, Atmanspacher, & Pothos, 2013).

We now take a brief, non-technical peek at how this research leads to a new conception of creativity, and in turn, a new conception of what drives culture. In the 'quantum' approach to concepts, a concept can exist in different states, and there is a relationship between the strengths, or *weights,* on the properties of a concept in a particular state, and the concept's susceptibility to change, or *collapse* to, any particular new state. For example, if you think about FIRE in terms of only its most typical properties such as 'hot', your next thought may be about something else that is hot, such as an OVEN. However, if you think about FIRE in a way that encompasses not just typical properties such as 'hot' but also atypical properties, and in particular those implied by the current context, your next thought may be about something semantically distant from FIRE; for example, a poet might think of a word that rhymes with FIRE such as HIRE. The state of the concept when you are not thinking about it—so there is no context influencing it—has been referred to as its *ground state*. In its ground state, there are no properties that are definitively included in the concept, but also, no properties definitively excluded from it. This means that, for any concept, there exists some context for which even a seemingly defining property of that concept could be excluded. An example is KITCHEN ISLAND, for which a seemingly defining property of ISLAND, namely 'surrounded by water,' is (hopefully!) excluded. Likewise, there exists some context that could come along and make any given property become relevant. For example, 'edible' is not usually a property of HOUSE, but it is a property of GINGERBREAD HOUSE. The more exotic the context, the more atypical the properties that are evoked, and thus, the more unconventional the subsequent thought. Much creative cultural change involves exploiting the potentiality of concepts to collapse different ways when they appear in different contexts.

Since understanding this is central to understanding the process that fuels the generation of cultural novelty, let's examine it in more detail using a metaphor that is illustrated in **Figure 9**. The figure depicts two woodcuttings with light shining on them from three different directions, yielding three differently shaped shadows: that of a G, an E, and a B. Consider the situation in which you cannot see the woodcutting, and must guess its shape by examining the shadows it casts. Three different shadows are projections of the same underlying object. You are in a state of uncertainty regarding its shape, but these shadows constrain the space of viable answers. In the language of the above-mentioned concept combination models, the state of the woodcutting when no light is shining on it is its *ground state*. The woodcutting has the potentiality to *actualize* different ways, and to actualize in one of these ways requires an *observable* or context, in this case, light shining from a particular direction.

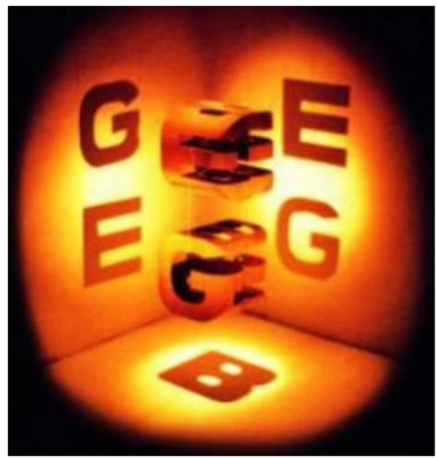

**Figure 9**. Photograph of ambiguous wood-cuttings taken from the front cover of (Hofstadter, 1979). The top `trip-let' (as the author refers to them) is not simply a rotated version of the one below it; it is a different shape. (Used with permission.)

The metaphor between creative cognition and these woodcuttings sheds light (so to speak) on many aspects of the creative process. Consider first the upper woodcutting. Your uncertainty as to its actual shape represents your uncertainty regarding what form the particular creative output you are working on will eventually take. The act of shining a light on the woodcutting from different angles represents the process of looking at the creative problem or task from different perspectives. The honing of your creative idea is represented by this process of reiteratively shining light on the woodcutting, first from one direction, then from another, until you glean the shape of the woodcutting that is hidden from you.

This metaphor suggests a way of understanding the creative process that conveys something of the flavour underlying quantum models of concept combination. Since the different sketches of a painting, or prototypes of an invention, take different forms when expressed in the physical world, it is tempting to assume that that they derive from different underlying conceptions in the mind. Thus, it seems self-evident that a stream of creative thought entails the generation of multiple distinct, separate ideas, and thus, that creativity amounts to a process of trial and error. However, much as that the three shadows of each wood-cutting are projections of the same underlying object, the above-mentioned quantum cognition research suggests that there may be a single underlying as-yet-poorly articulated vision of the idea that manifests as different sketch or prototype ideas when looked at from different perspectives (Gabora, 2019b; Scotney et al., 2020).

**The Creative Process**

Quantum cognition research led to a new conception of convergent and divergent thought (Gabora, 2019b). It suggests that in convergent thought an idea is refined by conceiving of concepts and ideas in their *conventional contexts*. Because one is not concerned about all the remote ways in which the object of thought could be related to other things, but instead working with it in its most compact form, mental energy is left over for complex operations. This then is why convergent thought is conducive to unearthing relationships of causation, or thinking analytically, as well as simply carrying out rote tasks.

Conversely, in divergent thought one reflects on an idea by considering a particular concept or idea from *unconventional contexts*. This is conducive to unearthing relationships of correlation, i.e., forging new connections between seemingly unrelated areas, as in analogical thinking. Note that the more unconventional the contexts one calls up, the seemingly less sensible the next thought may be, and therefore the more creative reflection, or *honing,* that may be required to coax it into a form that eventually makes sense. It is for this reason  because of the inherent ambiguity of distant connections compared to close ones that the products of divergent thought (as redefined here to mean thinking of ideas from unconventional contexts) may require extensive honing.

The woodcutting metaphor also provides a way of envisioning what happens as you hone or reflect on an idea. Midway through a creative process you may have an inkling of how to proceed, but not yet know whether, or exactly how, it could work; we say that the idea is `half-baked.' By thinking divergently, you entertain wildly different ways of looking at it, and these different perspectives yield wildly different 'takes' on the idea, much as shining light on the woodcutting from different directions yields different shadow-letters. However, over time, your reflections on the problem converge. You start thinking divergently, looking at it from perspectives that are less wide-ranging, and the images that come to mind overlap more. Your conceptions of it seem less like 'different ideas' and more like 'variants of one idea,' much as shining light on the upper woodcarving from the right at slightly different angles would yield similar shadows. Although some might be wobbly or distorted, they would all be G-shaped.

**The Guiding Role of Global Conceptual Structure**

It has long been said that the creative process is initiated by a problem, inconsistency, or unexpected finding, or a sense of fragmentation, curiosity, restlessness, sometimes been referred to as 'the gap.' The gap can be something as trivial as the sense that a little scribble lacks a sense of completion, or as profound as the need to solve a complex problem such as poverty or climate change. There is an increase in what Hirsh, Mar, and Peterson (2012) call *psychological entropy*: arousal-generating uncertainty, which can be experienced negatively as a source of anxiety, or positively as a *wellspring for creativity* (Gabora, 2017).[5]

Despite the notion that creativity begins with a 'gap,' there is little discussion in the creativity literature of what it is that *has* a gap. There is talk of a 'problem domain,' but since creative insights often come about through analogies, or the merging of ideas from different domains (Gentner, 1983; Holyoak & Thagard, 1996; Scotney, Schwartz, Carbert, Saab, & Gabora, 2020), the creative process is not confined to a problem domain; it builds bridges across one's worldview. Indeed, semantic network approaches to creativity show that the associative networks of highly creative people are more interconnected than those of less creative people (Kenett, 2018). Thus, the gap can be understood as a gap in the worldview: a fragment of understanding that does not yet cohere with one's web of understandings, which challenges us to spontaneously reorganize them, so as to make sense of the unexpected, or express something that needs to come out. It is because worldviews constitute a second level of self-organizing structure that people can channel arousal in creative directions: they reflect on the contents of their worldviews, revising and considering them from different perspectives, until psychological entropy reaches an acceptable level, and the creative output feels complete. Through play, exploration, imagination, fantasy, and hypothetical 'what if' type thinking, a worldview may reach a more coherent configuration.

The woodcutting metaphor provides a nice visual explanation of why we cannot understand the creative process without investigating the global structure of the creator's conceptual network. Note that the two wood-cuttings have different shapes, yet they yield the same three shadows. To distinguish the shape of the woodcutting above from the one below would require that light be shown on them from still more angles, casting shadows that would not look like any particular letters we know. Similarly, the more complex one's unborn creative idea, the more honing steps required to discern its underlying form. <NEXT BIT NOT QUITE RIGHT, BECAUSE THERE ARE TWO LEVELS HERE, REALITY ITSELF, AND YOUR CONCEPTION OF IT, AND THO WHO KNOWS THERE MAY BE AN IMPACT OF YOUR CREATIVE THOUGHT ON THE UNDERLYING REALITY ITSELF, IT IS THE CHANGE TO YOUR CONCEPTION OF THE IDEA IN THE HUMNAN REALITY THAT IS OF ISSUE HERE: Note also that the situation in creativity may be more complex than the woodcuttings scenario because each time you think about the creative problem from a particular angle, that thought may feed back and alter your conception of the problem. (It's as if each time you cast a shadow on a woodcarving, that slightly altered the shape of the woodcarving.) Moreover, there may be order effects; that is, the order of the perspectives from which the problem is looked at may alter the solution one finds or the creative output one ends up with. The upshot is that to understand the creative process we need to understand how and why the creator looks at the problem from a particular sequence of perspectives, in a particular order. In turn, to understand this, we must look beyond models of concept combination to the global structure of how a creator's concepts are woven together into a conceptual network, and how this network is restructured, fueled by the creator's needs, goals, and desires.

---

[5] Psychological entropy may be related to the notion that people constantly recalibrate their mental models on the basis of new sensory information to minimize free energy (Friston, 2010).

Thus, the creative process reflects the natural tendency of a worldview to seek a state of dynamic equilibrium by exploring perspectives and associations. A creative outcome (e.g., a painting) can be an *external* manifestation of *internal* cognitive restructuring brought about through immersion in a creative task. The creator may see and feel the world differently afterward, which is why creativity can be transformative, and why expressive art therapies are gaining prominence. Not all creative outputs involve extensive cognitive restructuring; some are minor variations on a theme, and others outright imitations, which is why a creative work cannot be fully understood outside of its cultural context.

When individuals work together in creative teams, they collectively possess a richer repertoire of perspectives from which to look at, and hone, the emerging idea. When one individual has taken the idea as far as they can, we can say it reaches an end state (or 'eigenstate') with respect to the current state of that person's current worldview. However, the same idea, when confronted by someone else's worldview, may seem unfinished; it may appear to be possible to further refine tie idea, or take it further. Team members build on each other's ideas by viewing them from different yet overlapping knowledge bases, and adapting them to their own taste, needs, and desires. Multicultural teams can be particularly creative, and indeed, by exposing people to a larger pool of ideas, multicultural experiences can have a stimulating effect on creative performance (Leung et al., 2008; Leung & Chiu, 2010).

<FIND PLACE FOR THIS> the inspirational sources for creative outputs often come from domains other than that of the creative output itself (Scotney, et al, 2019), and analogy and metaphor play a central role in creativity (Gentner, 1983; Holyoak & Thagard, 1996; Scotney et al, 2020).

## Honing Theory: Core Tenets

The core tenets of honing theory can thus be summarized as follows:

1. Formal research on concept combination suggests that *we experience a concept or idea only indirectly*, by observing how it appears when reflected upon from different perspectives. The *generative phase* of creativity involves, not generating multiple discrete, separate ideas, but generating multiple 'shadow-like' projections of the underlying problem (and its possible solution), by viewing it from different perspectives.

2. The creator wrestles with a single underlying mental construct that can (nonetheless) take different forms when expressed in our three-dimensional world. As the creative process proceeds, what is changing is not the number of ideas generated, but the diversity of perspectives the idea is looked at from. As one transitions from shining light on an object from different directions to shining it from almost-identical directions, what was initially multiple non-overlapping shadows coalesce into a single shadow. Similarly, as one transitions from reflect on an idea from different (divergent) perspectives to increasingly similar (convergent) perspectives, it starts to feel less like generating *many* ideas and more like refining a *single* idea.

3. *Engagement in the creative act transforms a creator's worldview*. This transformation extends beyond the 'problem domain' to affect the global structure of the worldview as a whole, which is is why immersion in a creative task can be therapeutic. It can establish or strengthen the creator's identity and personal style, or 'voice', and be accompanied by a sense of release (Barron, 1963; Forgeard, 2013), which is recognizable not just *within* domains but also *across* domains (Gabora, O'Connor, & Ranjan, 2012).

4. It is through the transformative restructuring of worldviews *that the creative process fuels culture*, and *culture evolves*.

## Question 3: How Does Human Culture Evolve?

Let us now turn to the third question put forward in the Introduction, that of how human culture evolves. Much human behavior, like that of other organisms, is biological in origin; we are driven to eat, mate, and do what it takes to survive and reproduce such that our genes live on. Yet we also hear the call of a second evolutionary imperative: that of culture. We learn cultural knowledge, carry out rituals, and obey social norms, and perhaps most intriguingly, we bring forth new seeds of cultural change. When people are in the throes of creative inspiration they may forget to eat, or ignore their children. They are acting in the service of, not biological needs, but cultural needs, behaving in ways that propagate not genetic lineages, but cultural lineages.

Thus, it appears that humans are propelled by a second kind of evolutionary force. We have as much to gain by establishing a scientific understanding of how culture evolves as we have accomplished through establishing a scientific understanding of how organisms evolve. A formal theory of culture evolves will amplify our understanding of who we are, where we come from, and where we may be headed. Indeed, we have even more to gain by developing a multidisciplinary framework for evolutionary processes that encompasses both biological and cultural evolution under the same umbrella.

We now address the question of how culture evolves.

## A Darwinian versus a Non-Darwinian Approach to Cultural Evolution

Early attempts to develop a scientific framework for culture began with the hypothesis that culture evolves, as does biological life, through a Darwinian process, and Darwinian theories of cultural evolution have been around for some time (Boyd & Richerson, 1985; Cavali-Sforza & Feldman, 1981; Gabora, 1997).

### Acquired versus Inherited Traits

Darwin's theory arose in response to the paradox of how organisms accumulate adaptive change despite that traits acquired over a lifetime—referred to as *acquired traits*—are eliminated at the end of each generation.

In some domains that exhibit cumulative, adaptive change—e.g., cultural evolution, and primitive forms of life—acquired traits are not always extinguished at the end of a generation; therefore, the paradox that Darwin's theory was designed to solve does not exist! When you think about it, this lack of transmission of acquired traits across biological lineages is striking and unique. If an asteroid goes from being jagged to rounded as it moves through space, it doesn't generally revert back to possessing the trait of being jagged. Once cups had handles, they didn't revert back to the state of not being able to have a handle. But changes you acquire over your lifetime—such as a tattoo, or knowledge of the layout of your town—are not passed on to your children, and thus a biological lineage (such as from grandparent to parent to child) reverts back to an earlier state at the beginning of each generation. Why is that?

*Darwin's theory of natural selection* posits that (1) acquired traits (those an organism picks up over its lifetime) are discarded (e.g., you don't inherit your mother's tattoo), but inherited traits (those passed down through the genes) are retained (e.g., you *can* inherit your mother's blue eyes), and (2) inherited traits that enhance fitness are selected, such that over generations, they become more widely represented in a population. Advocates' rationale for the approach is that the algorithmic structure of natural selection is not dependent upon its biological substrate, and indeed this is the case; genetic algorithms successfully abstract this structure and apply it to optimization (Holland, 1992).

*Random Variation*

Darwin's theory is only applicable to a process to the extent that the source of novelty, or variation, is randomly generated, which is why population geneticists who developed a methmatical framework for natural selection made 'random variation' a precondition for the apopriateness of a Darwinian framework (Fisher, 1930; Hartl & Clark, 2006). To the extent that novelty is *not* randomly generated, change over time is due to whatever causal agent is biasing the variation away from random, not to selection. In other words, it is due to whatever is making some come into existence in the first place and not others, as opposed to a pruning down of those already in existence. Cognitive processes generate variation *non-randomly,* which violates one of the assumptions that must be in place for Darwin's theory of natural selection to be applicable. This is not the only reason a Darwinian framework for culture breaks down, as we will see, but it is a sign that the Darwinian approach to culture is on the wrong track.

Nonrandom creative processes such as analogy can bring about adaptive change instantaneously, rather than requiring generations to exert an effect, as is the case for processes that evolve through natural selection. A process that exerts change instantaneously can quickly drown out any impact of one that requires generations to exert any effect; it will "swamp the phylogenetic signal."

In addition, cultural entities of very different kinds 'mate' with each other; shopping malls are full of things like Mickey Mouse shaped alarm clocks, Scrabble ledge inspired furniture, or even dumbells that function as water bottles. In biological evolution, matings between vastly different kinds (such as betweeen different species) are detrimental and often lethal, because their genetic instructions (DNA) are incompatible.

Given these salient differences, does the cultural evolution of behavior and artifacts really have the same algorithmic structure as the biological evolution of species? To answer this question, we must examine the theory of natural selection in a little more detail.

The lack of transmission of acquired traits comes about because of the sequestering (or 'hiding away') of germ cells from environmental change (i.e., change acquired during the lifetime).[6] This, in turn, requires a *self-assembly code* that is used in two distinct ways (Holland, 1992; von Neumann, 1966). First, it is actively interpreted during development to generate a body, or soma. Second, it is passively copied without interpretation during reproduction to generate germ cells (such as sperm and eggs).

---

[6] Keep in mind that this is what is necessary for Darwinian theory to hold; i.e., for a process to be explain

However, cultural evolution does not involve a self-assembly code used in these two distinct ways. Since there is no sequestration of germ cells from developmental change, acquired changes *can* be transmitted, and therefore a lineage doesn't periodically revert back to an earlier state as is the case for evolution through natural selection.

It has been proposed that cumulative, adaptive change in both culture, and the very earliest life forms on earth, reflects a lower-fidelity non-Darwinian evolutionary process (Gabora, 2004, 2008b, 2013, 2019; Gabora & Steel, 2021). We refer to this more primitive evolutionary process as *Self-Other Reorganisation* (SOR) because it involves internal self-organising and self-maintaining processes within entities, as well as interaction between entities, as illustrated in **Figure 9**. SOR encompasses learning, but in general operates across groups. It differs from what is called neutral evolution in that it entails adaptive change, albeit it is less efficient than evolution through natural selection.

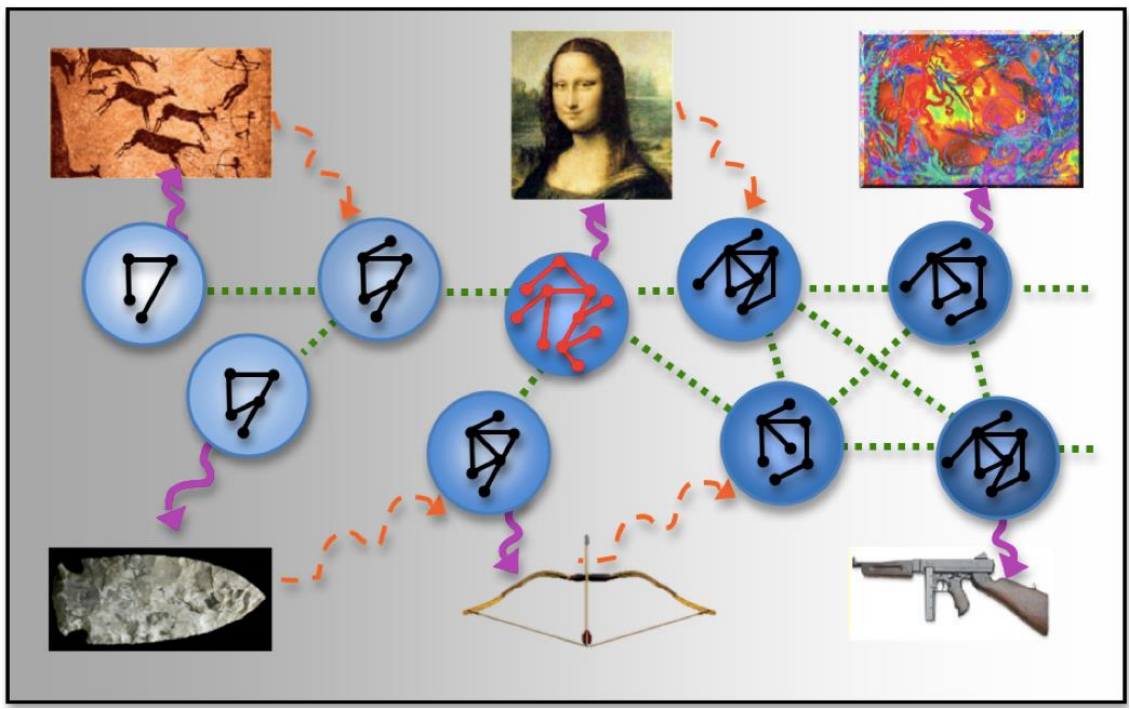

**Figure 9.** Schematic depiction of SOR operating in a cultural context, showing how worldviews evolve through, not *survival of the fittest,* but *transformation of all. Worldviews* transform as a consequence of psychological entropy-reducing restructuring and communal exchange. Individuals are represented by spheres and their internal models of the world, or *worldviews,* are represented by networks within the spheres. Patterns of social transmission are indicated by dashed green lines. Creative contributions to culture are indicated by wavy purple arrows from creator to artifact. Learning through exposure to artifacts is indicated by wavy orange arrows. Worldviews and patterns of social transmission tend to become more complex over time. Individuals such as the one with the red network are more compelled than others to reframe what they learn in their own terms, potentially resulting in a more unique or nuanced worldview. Such 'self-made' individuals are more likely to have an effect on culture (*e.g.,* through the creation of artifacts), which may influence the formation of new worldviews long after they have died. This is indicated by the diffusion of segments from self-made individuals to other individuals, either indirectly by way of exposure to the Mona Lisa, or directly by way of exposure to its creator. From (Gabora, 2013).

In culture, the equivalent of an organism—i.e., what is evolving through SOC—is our worldviews (i.e., our minds as they are experienced from the inside). To understand how a particular individual comes up with an idea, and what other members of that individual *do* with that idea, you have to know something about how different abilities, such as learning, memory, decision making, and emotion, function together as an integrated whole, i.e., a worldview. You can think of a worldview as having a particular shape that reflects the individual's basic orientation to life: how they see things as being related to each other, how they interpret situations, what they tend to dwell upon or avoid, as well as how complex and dynamic their understanding of the world is. An individual's behavior, how they creatively express themselves, and how they respond to situations, reflects (to some extent) the unique architecture of their worldview. Since knowledge exhibits intermediate modularity, consisting of loosely connected clusters that can be characterized using tools from network science, network science offers a promising avenue for a formal process model of the emergence and organization of conceptual structure, including individual differences, and the role of arousal and emotion on conceptual structure (Kenett, 2018).

Network models of cognition unleash a new psychologically valid approach to understanding how civilizations grow and evolve, and organizing cultural innovations into cultural lineages. Conventional approaches to organizing artifacts into chronological lineages a assume Darwinian framework, and categorize them solely on the basis of predefined superficial attributes (e.g., whether a jug has a handle or not). In contrast, in the SOR approach, artifacts are classified not just on the basis of objective superficial attributes but also on the basis of what makes them *meaningful to humans,* including symbolic and utilitarian functions and affordances (e.g., despite dissimilar physical attributes, PESTLE and MORTAR are conceptually close because together they perform a single task), as well as the underlying cognitive processes required to transition from one artifact to another (e.g., concept combination, or analogy) . In analyses of Baltic psalteries and PaleoIndian projectile points (which both have a well-documented ethnographic and archaeological record) the network approach recovered previously unacknowledged patterns of historical relationship that were more congruent with geographical distribution and temporal data than what was obtained with other approaches (Gabora et al., 2011; Veloz et al., 2012).

## Cognitive Autocatalytic Networks: The Hub of Cultural Evolution

Our lab has been using a certain kind of abstract network to model human cognition, referred to as ***autocatalytic network*** (Kauffman, 1986, 1993). The term *autocatalytic* refers to the fact that interactions amongst initially distinct *parts* can give rise to (i.e., *auto*nomously *catalyze* the emergence of) a new *whole*. Autocatalytic network theory grew out of studies of the statistical properties of random graphs consisting of nodes randomly connected by edges. As the ratio of edges to nodes increases, connected nodes join to form clusters, and as the size of the largest cluster increases, so does the probability of a phase transition resulting in a single giant connected cluster, i.e., an integrated whole. The point at which this happens is referred to as *percolation threshold*. Thus, autocatalytic networks were used to develop the hypothesis that life began as, not as a single self-replicating molecule, but a set of molecules that, through catalyzed reactions, collectively reconstituted the whole (Kauffman, 1993).

The autocatalytic approach offers an established formal framework for modeling the emergence of structures that evolve through SOR. Unlike other network science approaches, autocatalytic models focus not just on network structure per se, but on how the network acquires the capacity to spontaneously reconfigure itself in response to current needs. They have been applied extensively to not just biological evolution (Hordijk, Hein & Steel, 2010; Hordijk, Kauffman & Steel, 2011; Hordijk & Steel, 2013; Kauffman, 1986, 1993; Steel, 2000; Steel, Hordijk, & Xavier, 2019; Steel, Xavier, & Huson, 2020; Xavier, et al., 2020) but also cultural evolution (Gabora, 1998, 1999; Gabora & Steel, 2017, 2020a, 2020b, 2022; Ganesh & Gabora, 2022). Autocatalytic networks apply equally well to the study of biological or cultural networks because they are simply an abstract mathematical framework, with multiple applications. Indeed, it has been proposed that they may hold the key to understanding the origins of *any* evolutionary process (Gabora & Steel, 2021). This approach thereby frames the study of cultural evolution within the overarching scientific enterprise of understanding how evolutionary processes (be they biological or cultural) begin, and unfold over time.

In applications of autocatalytic networks to culture, the mind plays the role in cultural evolution that the soma plays in biological evolution. The observation that, like living organisms, cognitive networks are self-sustaining, self-organizing, and self-reproducing (Maturana & Varela, 1973; Varela, Thompson, & Rosch, 1991) suggests that cognitive networks constitute a second level of autocatalytic structure.[7] The self-sustaining nature of a cognitive network is evident in the tendency to reduce cognitive dissonance, resolve inconsistencies, and preserve existing schemas in the face of new information. Although the contents of a cognitive network change over time, it maintains integrity as a relatively coherent whole. Its spontaneously self-organizing nature is evident in the capacity to combine remote associates (Mednick, 1962) (such as combining LIPS and STICK to invent LIPSTICK). It reproduces in a piecemeal manner through imitation and other forms of social learning, as well as the propensity to share stories and perspectives.

The origin of culture involves, not chemical reaction networks, but networks of knowledge and memories, and the products and reactants are not catalytic molecules but mental representations. Thus, in cognitive applications of autocatalytic networks, the nodes represent mental representations, including memories, chunks of procedural or declarative information, scripts, and schemas. Mental representations may be composed of one or more *concepts:* mental constructs such as CAT or FREEDOM that enable us to interpret new situations in terms of similar previous ones. Just as reactions between molecules generate new molecules, interactions between mental representations generate new mental representations, which, in turn, enable new interactions. Just as a catalyst makes a certain reaction more likely to occur, a question, desire, or external stimulus can trigger or 'catalyze' a new idea, or perspective.

In the case of cognitive networks, the foodset consists of *existing information* that is either innate, or obtained through individual learning or social learning. Thus, for example, learning from personal experience how to distinguish a particular bird from other birds would result in a foodset item, as would learning from a teacher the scientific name for the bird. A foodset-derived item consists of *new information* generated by you. For example, a poem or song that you write about the bird, or a theory you develop about its role in the ecosystem, would be a foodset-derived item. (In reality, the distinction between foodset and foodset-derived items is not always black and white, but we can ignore that subtlety for now.) Whereas conventional 'node-and-edge' network models require external input to continue processing, autocatalytic networks 'catalyze' conceptual change endogenously, resulting in new conduits by which goals can be met. An autocatalytic network is self-organizing, and conceptual change can percolate throughout the network and affect its global structure.

---

[7] By *cognitive network,* we refer to an individual's web of concepts, language terms, and their associations, as well as knowledge and memories, and how they are structured.

The process by which a new foodset-derived item (e.g., the song, poem, or theory) is generated out of foodset items is referred to as an ***interaction*** or ***reaction.*** A key notion in cultural RAF theory is that an interaction may be facilitated by a particular stimulus, drive, or thought. For example, you *might* just happen to write a poem about lasagna, but it is more likely that this would be triggered or *catalyzed* by a stimulus such as seeing or smelling lasagna, or a memory of a particular instance of lasagna, or the drive state of hunger. The poem is foodset-derived because it is composed of foodset items (words and their referents) that are put together in a novel way, resulting in something new. The ***reactivity*** of a concept or idea refers to its capacity to stimulate or catalyze recursive sequences of conceptual change. Ideas may become highly reactive by aligning with needs, drives or desires.

In the above examples, the foodset-derived items were more complex than the foodset items of which they were composed, but foodset-derived items may also be simpler; for example, a new foodset-derived item might come about through the realization that two seemingly distinct entities are both instances of the same more abstract kind of entity. Thus, the mental representations in such a mind are bound together by a web of associations, and relationships of logic, causation, and so forth. It is proposed that these interconnections enable such a mind to adapt existing protocols to new situations, combine concepts in new ways, and put their own spin on ideas such that they reflect their own preferences and emotional states, which in turn enabled cumulative cultural evolution.

*Reflexively Autocatalytic Foodset Generated (RAF) networks*

Autocatalytic networks have been mathematically extended and generalized as ***Reflexively Autocatalytic Foodset Generated (RAF) networks*** which are more broadly applicable beyond their original application to the origin of life (Steel, 2000). RAFs are the form of autocatalytic networks we find most useful in modeling cultural evolution. The term ***reflexively*** is used in its mathematical sense, meaning that each part is related to the whole. The term ***foodset*** refers to the elements that are initially present, as opposed to ***foodset-derived*** elements, which are the products of interactions amongst them. Technically, a RAF is thus a non-empty subset of 'reactions' that satisfies the following properties:

1. *Reflexively autocatalytic:* each reaction is catalyzed by at least one element type that is either produced by the network or present in the foodset.
2. *F-generated:* all reactants can be generated from the foodset through a series of reactions in the network.

The collection of all RAFs in a system forms a partially ordered set (a poset) under set inclusion, with a unique maximal element, referred to as the *maxRAF*. RAFs can be hierarchically structured, i.e., the maxRAF may consist of multiple levels of subRAFs. The distinction between foodset and foodset-derived items enables us to study under what conditions these (pre-existing + newly generated) items collectively become integrated wholes of various types.

Following up on Kauffman's (1993) above-mentioned early application of autocatalytic networks to the origin of life, significant progress has been made using RAFs to model how interactions amongst chemicals present in earth's early atmosphere could have given rise to a set of chemicals that collectively replicates and evolves, i.e., a living protocell (Steel, 2000; Hordijk, Hein, & Steel, 2010; Hordijk, Steel, Kauffman, 2012; Hordijk, Hasenclever, Gao, Mincheva, & Hein, 2014; Hordijk, 2016; Xavier, Hordijk, Kauffman, Steel, & Martin, 2020). Autocatalytic network theory has successfully demonstrated—mathematically or using simulations (Hordijk, Hein & Steel, 2010; Hordijk, Kauffman & Steel, 2011), and with real biochemical systems (Hordijk & Steel, 2013; Xavier, et al., 2020)—how self-maintaining structures that evolve and replicate (through amplification and drift acting on possible subRAFs of the maxRAF) can emerge from nonliving molecules.

While in those studies RAFs were used to model the emergence of *biologically* evolving structure, here we focus on the use of RAFs to model the emergence of *culturally* evolving structure. Specifically, RAFs are used to model the transition from the kind of mind that consists primarily of mental representations that we categorize as foodset items (obtained by learning from the environment, or social learning) to the kind of mind that generates sufficient foodset-derived mental representations to unite these foodset mental representations into an integrated web: a conceptual network, or worldview (**Figure 10**). Thus, RAFs are useful for modeling the kind of cognitive structure capable of generating cumulative, adaptive, open-ended innovation (Gabora & Steel, 2017, 2020a, 2020b, 2021, in press; Gabora, Beckage & Steel, 2022; Ganesh & Gabora, 2022).

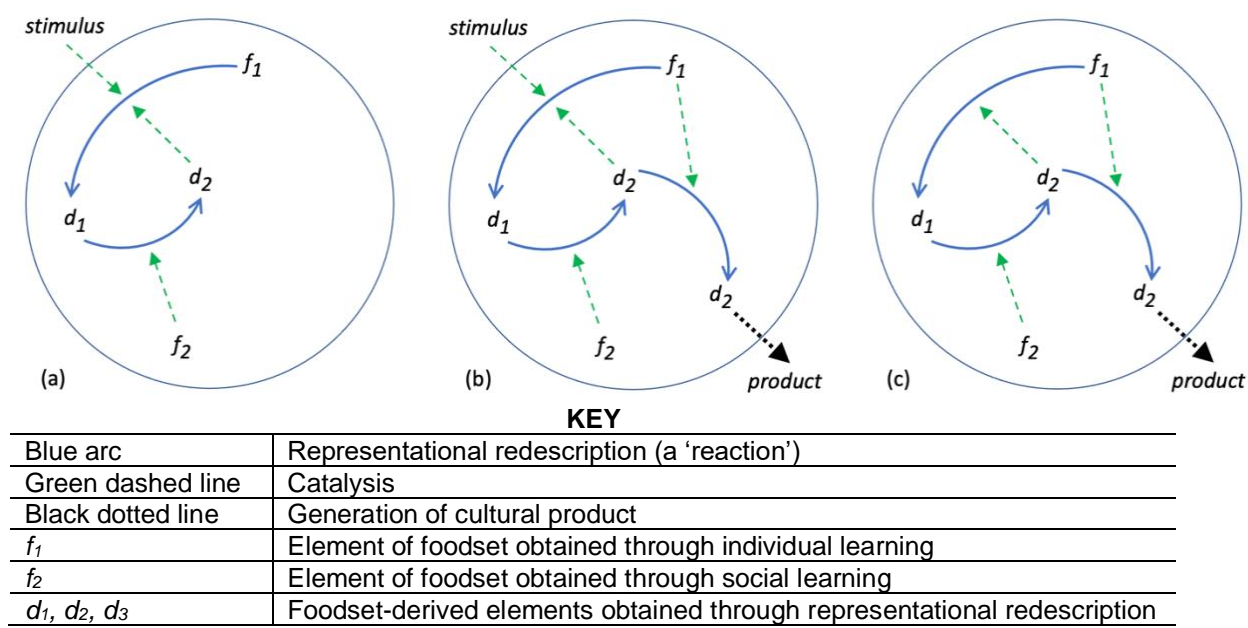

**KEY**

| Blue arc | Representational redescription (a 'reaction') |
|---|---|
| Green dashed line | Catalysis |
| Black dotted line | Generation of cultural product |
| $f_1$ | Element of foodset obtained through individual learning |
| $f_2$ | Element of foodset obtained through social learning |
| $d_1, d_2, d_3$ | Foodset-derived elements obtained through representational redescription |

**Figure 10**. Emergence of a simple RAF network. (a) A stimulus sets in motion a sequence of representational redescription events, resulting in creative transformation of a foodset item obtained through individual learning. (b) This culminates in a new product. (c) Since the first representational redescription event can be catalyzed not just by a particular stimulus but also by a derived element, once the chain of events once set in motion by the stimulus, it is no longer dependent on the stimulus. Therefore, it is referred to as a persistent RAF, as opposed to a transient RAF.

*Advantages of the RAF Approach*

*Modeling Endogenous Change*. What differentiates autocatalytic networks from other network science approaches is that nodes are not just passive transmitters of activation; they actively galvanize, or 'catalyze' the synthesis of novel ('foodset-derived') nodes from existing ones (the 'foodset') (Gabora, Beckage & Steel, 2022). Because RAF nodes modify network structure, the RAF framework is consistent with the goal of understanding not just how networks are structured, but also how they dynamically restructure themselves in response to internal and external pressures. This makes RAFs uniquely suited to model how new structure grows out of earlier structure, i.e., generative network growth (Steel, 2019). Such generativity may result in phase transitions to a network that is self-sustaining and self-organizing (Hordijk & Steel, 2004; Hordijk, 2016; Mossel & Steel, 2005), as well as potentially able to self-replicate (in a relatively haphazard manner, without reliance on a self-assembly code), and evolve, i.e., exhibit cumulative, adaptive, open-ended change (Gabora & Steel, 2021; Hordijk & Steel, 2015).

*Potential to Scale Up*. The generalized RAF setting is conducive to the development of efficient (polynomial-time) algorithms for questions that are computationally intractable (NP-hard) (Steel, Hordijk, & Xavier, 2019).

*Semantic Grounding*. When applied to cognition, the demarcation into foodset versus foodset-derived elements provides a natural means of (i) *grounding abstract concepts in direct experiences* (foodset-derived elements emerge through 'reactions' that can be traced back to foodset items), and (ii) precisely describing and tracking how new ideas emerge from earlier ones. Thus, RAFs can model how endogenous conceptual restructuring results in new conduits by which uncertainties can be resolved, ideas can be generated, and needs can be met. A source of uncertainty is modeled as an element that resists integration into the individual's cognitive network, or worldview, which is described as a maxRAF containing the bulk of the individual's mental representations. This disconnect produces arousal, which catalyzes one or more interactions amongst mental representations. A single conceptual change can precipitate a cascade of reiterated cognitive 'reactions' (self-organized criticality) that affect the global structure of the maxRAF.

## RAFs applied to the Origin of Culture

As mentioned earlier, we proposed that the transition from Oldowan to Acheulean tool technology 1.76 mya was precipitated by the onset of representational redescription, which enabled the forging of associations, and the emergence of hierarchically structured concepts, making it possible to shift between levels of abstraction to carry out complex tasks (Gabora & Smith, 2018). Unlike more primitive Oldowan stone tools, the Acheulian hand axe required, not only the capacity to envision and bring into being something that did not yet exist, but hierarchically structured thought and action, and the generation of new mental representations: the concepts EDGING, THINNING, SHAPING, and a meta-concept, ACHEULIAN TOOL. We identified what socially or individual learned 'raw ingredients' were needed for these concepts, as well as the drive state that catalyzed their transformation (through representational redescription) into the new concept. In our model, this resulted in *transient RAFs:* RAFs that rely on, and exist only in the presence of a particular external stimulus or internal drive (i.e., a specific 'catalyst'). This constituted a critical step toward what has been referred to as conceptual closure, characterized by the emergence of persistent autocatalytic cognitive structure.

The approach provides a promising approach to unraveling one of the greatest archaeological mysteries: that of why development of the Acheulian hand axe was followed by over a million years of cultural stasis. Our model suggests that the rate of representational redescription did not rise above the critical percolation threshold, so streams of abstract thought were short, and few in number.

## RAFs applied to the Origin of Behavioral and Cognitive Modernity

As mentioned earlier, the Paleolithic marked the appearance of art, science, and religion, and people who were behaviorally and cognitively 'modern,' i.e., they thought and felt and lived their lives in a way that feels recognizably human. Synthesizing research from psychology, anthropology, and cognitive science, it was suggested that this was brought on by one or more genetic mutations that allowed thought to be spontaneously tailored to the situation by modulating the degree of (1) divergence (versus convergence), (2) abstractness (versus concreteness), and (3) context-specificity. This in turn marked a phase transition culminating in the emergence of persistent, unified autocatalytic RAF networks that bridged previously compartmentalized knowledge and experience, and collectively formed a hierarchical maxRAF (Gabora & Steel, 2020b). Once this was in place, a problem, described as a point disconnected from the maxRAF, could be reflected upon from different perspectives, until this resulted in a solution, which is described by a subRAF connecting it to the maxRAF. This, in turn, marked the onset of the capacity to creatively adapt ideas to ever-new situations in a genuinely human way, and the beginnings of cumulative, adaptive, open-ended cultural evolution.

*Using RAFs to Construct Cultural Lineages*
In the cultural evolution literature, it is commonly assumed that two processes contribute to cultural evolution: social learning and individual learning. Creative thought gets lumped in with individual learning, but there is an important distinction between them. In individual learning (obtaining pre-existing information from the environment by non-social means through direct perception), the information does not change form just because the individual now knows it. Noticing for oneself that lightning tends to be followed by thunder is an example of individual learning. In contrast, in abstract thought (processing of internally sourced mental contents) the information is in flux, and when this incremental honing process results in the generation of new and useful or pleasing ideas, behaviour, or artifacts, it is said to be creative.

As we have seen, the RAF approach tags mental representations with their source, i.e., whether they were (1) innate, (2) acquired through social learning or individual learning (of *pre-existing* information, i.e., elements of the foodset), or (3) the result of creative thought (resulting in *new* information, i.e., foodset-derived elements). This demarcation makes it possible to distinguish the contribution of these three different sources on the emerging conceptual networks of individuals and social groups. It also makes it possible to trace innovations back to the individuals that generated them, precisely describe and track how new ideas and cultural outputs emerge from previous ones, and track cumulative change in cultural lineages step by step. RAFs can accommodate cultural discontinuities, which present a formidable challenge for models of cultural evolution (Kolodny, Creanza, & Feldman, 2015). The mechanisms by which psychological processes give rise to cultural outputs are not always evident from examination of the cultural outputs themselves (Heyes, 2018, 2019; Osiurak & Reynaud, 2020). Therefore, cultural lineages cannot be documented merely through the analysis of cultural outputs, and a theory of cultural evolution must incorporate the conceptual networks and psychological processes that spawn cultural innovations. Cultural discontinuities have their origin in cognitive processes such as analogy and cross-domain inspiration, as well as social, emotional, and personality factors that constrain and enable the exchange of ideas. The RAF approach can model the collaborative creativity and cultural discontinuities as intermittent bursts and cascades of cognitive 'reactions' within and between individuals (Gabora & Steel, in press; Ganesh & Gabora, 2022).

*RAFs applied to Cognitive Development*
RAF networks have been used to develop a step-by-step process model of conceptual change, *i.e.,* the process by which a child becomes an active participant in cultural evolution (Gabora & Beckage, 2022). At approximately age seven, children's thinking is primarily constrained by their everyday experiences, and this is modeled using a RAF comprised primarily of foodset items. By approximately nine years of age, children attempt to reconcile any inconsistencies between individually learnt and socially learnt information, and this is modeled as fragmentation of their maxRAF describing the conceptual network into discrete subRAFs. In the attempt to resolve these inconsistencies, children generate, test, and revise their understandings through abstract thought, resulting in new foodset-derived mental representations, modeled as the bridging of the subRAFs such that they are subsumed by the maxRAF. Abstract thought aids in the increasing coordination and hierarchically structuring of thought and action.

This was made more concrete using data on on childrens' mental models of the shape of the earth (Vosniadou & Brewer, 1992). We modeled different trajectories from the flat earth model to the spherical earth model, and modeled the impact of factors such as pretend play on cognitive development. As RAFs increase in size and number, they begin to merge and form a maxRAF that bridges previously compartmentalized knowledge. The expanding maxRAF constrains and enables the scaffolding of new conceptual structure. Once most conceptual structure is subsumed by the maxRAF, the child can reliably frame new knowledge and experiences in terms of previous knowledge and experiences, and engage in recursive representational redescription, or abstract thought, and adapt existing ideas to their own perspectives.

We suggested that individual differences in reliance on foodset versus foodset-derived information sources culminates in different kinds of conceptual networks and concomitant learning trajectories, which ultimately affects the kind and magnitude of the indivual's impact on cultural evolution. These differences may be amplified by differences in the proclivity to spontaneously tailor one's mode of thought to the situation one is in by modulating the degree of divergence (versus convergence), abstractness (versus concreteness), and context-specificity. More speculatively, this work could play a role in achieving 'general artificial intelligence,' by endowing computers with the human-like capacity to reflect on a problem by considering it from different perspectives.

*RAF Models of Diads and Teams*
I mentioned earlier that when individuals work together in creative teams, they collectively have a richer repertoire of perspectives from which to look at, and hone, the emerging idea. The RAF framework has been used to model the emergence of new conceptual structure through dyadic interaction. Interestingly, RAF structure can actually extend *across* individuals (Gabora, Beckage, & Steel, 2022; Gabora & in press).

## Future Directions

A next step in this line of research direction is to scale up the computational models of cultural evolution (MAV and EVOC) to model large societies using a massively parallel supercomputer, and replace the simple neural network based agents with large-scale computational cognitive RAFs.

There is also much to be done to better understand the creative processes that fuel cultural change. There are hints that interference and entanglement in concept interactions are related to creativity, but to date this has not been systematically explored. A related direction for future research concerns the role of incubation in creativity: the idea that setting a creative task aside for a while—i.e., incubating on it—can foster insight. One could model this as letting the idea return to its ground state such that it sheds its coterie of typical properties (and contexts), and taps into its reservoir of infinite potentiality (in the sense that no properties are definitively present nor absent). Merging research on creativity and conceptual interactions with RAF networks could result in more sophisticated cognitive RAFs that include the internal structure of mental representations and contexts. Moreover, the RAF approach could be used to model the emergence and honing of creative ideas in large teams of simultaneously interacting individuals, as a complementary approach to human studies of collaboration.

Another intriguing question worthy of future research is the following. Creative people are subject to both adoration social disapproval (and even bullying), and it is often assumed that this is because they violate social norms (Sternberg & Lubart, 1995). However, the notion that creative people engage in the process of honing ambiguous mental forms—which is consistent with evidence that they are comfortable with ambiguity (e.g., Tegano, 1990; but see also, Merrotsy, 2013)—suggests this may not be the whole story. Their comfort with ambiguity may extend to themselves, i.e., it may include ambiguity with respect to how they portray themselves and come across to others, which in turn may make them more vulnerable to other people's projections. In other words, they may be more subject to misinterpretation, appearing as devils to some, and as Gods or Goddesses to others.

The RAF model also lends itself to testable hypotheses. Because innate knowledge is biologically 'expensive,' it may be subject to ongoing evolutionary tinkering (i.e., a high mutation rate), to assess whether it continues to 'earn its keep.' This leads to the speculative prediction that a mutation that affects innate knowledge (modeled in the RAF approach as foodset items) may stimulate one to develop RAFs that replace innate sources of knowledge in others, thereby rendering it explicit, thus more amenable to analysis and modification. This is consistent with findings that intellectual giftedness can co-occur with learning disorders (Toffalini, Pezzuti, & Cornoldi, 2017). (Indeed, Einstein famously said that it was because he lacked the intuitive understanding of space and time that came naturally to others, he was forced to acquire a deeper understanding of these concepts (Einstein, 1949).)

The development of methods for identifying what information children have prior to the study (foodset items), assessing what kind of restructuring operations they are capable of, and determining when children generate new ideas (foodset-derived items), will facilitate future developmental applications of RAFs. This leads to another question that could be tested with RAFs: is someone who acquires a particular piece of information by learning it for themselves (i.e., in their mind it is a foodset-derived item) more likely to generalize that representation to other types of abstract models than someone who acquires it through social learning (i.e., in their mind it is a foodset item)? This hypothesis is reminiscent of those explored in developmental studies of generalization and causal reasoning (Gopnik, 2000, 2001). The RAF model offers an explanation of why cognitive differences might arise when learning is through instruction or imitation versus when learning is through opportunities to experiment and explore, and direct challenges that prompt students to learn these causal relationships for themselves. Other directions for future research include investigating whether RAFs could be useful for identifying gaps in knowledge, and exploring how cognitive development, social environment, and the form and content of socially transmitted information impact RAF formation. (For example, narratives that spread quickly through cognitive contagion may be describable as more compact RAFs.)

## Conclusions

During times of rapid change such as what we are living now, we have less to lose, and more to gain, from innovation. Existing rules and methods quickly become outdated, and holding onto them can be disastrous. New perspectives and ideas are needed to respond to unforeseen situations and glimpse new opportunities. Understanding how we innovate, how new ideas build on old ones, how the capacity for culture evolved, and in what sense human culture evolves, provides a conceptual framework for tackling the challenges of a changing world. The aim of the research program discussed in this chapter is to develop an integrated theory of the origin and evolution of the kind of cognitive structure that has culminated in humanity's unmistakable mark on this planet.

A key theme of this research program is that creativity and cultural evolution go hand in hand; you cannot have a complete theory of one without the other playing a central role. We saw that memory is distributed and content-addressable, and the more widely distributed mental representations became, the more they could overlap and evoke one another. Synthesizing evidence from cognitive psychology, neuroscience, anthropology, and archaeology, we posited that this was what gave rise to the origin of cumulative culture. Next we looked at the hypothesis that the origins of behavioral and cognitive modernity was due to onset of contextual focus: the capacity to shift between different modes of thought. In essence, these changes enabled us to forge and hone creative ideas, and it is through the 'honing' of creative ideas that worldviews transform and evolve. Support for these theories comes from testing the predictions they generate using computational models.

The chapter argued that cultural evolution required the emergence of cognitive structure that is self-organizing and self-reproducing, and evolves. Of course, minds are part of living organisms, which have these properties, but the focus here is the emergence of a second-order level of self-organizing structure that pertains not to cellular or organismal processes but to the webs of associations by which hominids weave together an understanding of their world. The raw materials for creativity come in the form of knowledge and experience in peoples' mental models, or worldviews; the creative process restructures them into new ideas, which take root and flourish, and transform as they spread to others. I propose that emergence of a new form of self-organizing, self-reproducing structure was as crucial to the origins of cultural evolution as it was to the origins of biological evolution, and the RAF network approach is particularly useful for modeling this. First, it distinguishes semantic structure arising through social or individual learning (modeled as foodset items) from semantic structure that is newly generated from this pre-existing material (modeled as foodset-derived items). Second, it accommodates the 'catalyzing' role of drives, emotions, and external stimuli in cognitive restructuring. These features of RAFs enable us to model how cultural novelty arises and integrated structure emerges, and to trace cumulative cultural lineages step-by-step. Although the RAF approach to cultural evolution is in its infancy, I believe it has the potential to provide a new framework for understanding historical change, one that offers a new and parsimonious explanation of its relationship to biological evolution, and that places the underlying cognitive processes centre stage.

## Appendix A

*Agent-based models* are computer programs that enable us to simulate the actions and interactions of autonomous agents in order to better understand their collective behavior as a whole. The agents can represent either individuals, or collective entities such as groups or organizations. In some agent-based models, the agents interact in an artificial environment; in others, their environment consists only of the other agents. Agent-based models enable us to manipulate specific parameters (such as the population size, population density, or invention-to-imitation ratio), so as to understand how these parameters affect the dynamics of the system and its outcomes in a more controlled manner than is possible in real life. For more information, see (Raczynski, 2020; Railsback & Grimm, 2019). For more information on their application to cultural evolution, see (Gabora, 2008, 2013; Gabora & Tseng, 2017).

The study of *autocatalytic networks* is a branch of network science, the study of complex networks (such as cognitive networks, semantic networks, social networks, or computer networks). Network science has roots in graph theory (e.g., Erdos & Rényi, 1960); elements of the network are represented by *nodes* (or *vertices*), and the connections between them are represented by *links* (or *edges*). Autocatalytic networks were originally developed to model the phase transition underlying the origin of life (Kauffman, 1986, 1993), but following their mathematical generalization to *Reflexively Autocatalytic Foodset Generated (RAF) networks* (Hordijk & Steel, 2015; Steel, 2000), have been applied to the origin of cultural evolution (Gabora & Steel, 2017, 2020a,b, 2021; Ganesh & Gabora, 2022), and complex cognition (Gabora, Beckage & Steel, 2022; Gabora & Steel, in press). What differentiates autocatalytic networks from other network science approaches is that nodes are not just passive transmitters of activation; they actively galvanize, or 'catalyze' the synthesis of novel ('foodset-derived') nodes from existing ones (the 'foodset') (Gabora, Beckage & Steel, 2022). (More information on autocatalytic networks can be found in the section of this chapter titled 'Cognitive Autocatalytic Networks: The Hub of Cultural Evolution').

## Acknowledgements

A huge thank you to Kirthana Ganesh, and particularly Isabel Gomez, for significant help with this manuscript. This research was funded in part by Grant 62R06523 to LG from the Natural Sciences and Engineering Research Council of Canada, and a gift to LG from Susan and Jacques Leblanc for research on creativity.